\documentclass[zpreprint,zbstdefault]{./zeus_paper}
\usepackage[english]{babel}
\newcommand{\ZcoosysA}{%
The ZEUS coordinate system is a right-handed Cartesian system, with the $Z$
axis pointing in the proton beam direction, referred to as the ``forward
direction'', and the $X$ axis pointing left towards the center of HERA.
The coordinate origin is at the nominal interaction point.\xspace}

\newcommand{\ZcoosysfnA}{\footnote{\ZcoosysA}}

\newcommand{\DO}{{D{\O}}\xspace}
\chardef\usc=95
\chardef\til=126
\catcode`\@=11 % @ signs are now treated as letters
\DeclareRobustCommand\xdotspace{\futurelet\@let@token\@xdotspace}
\def\@xdotspace{%
  \ifx\@let@token.\else
  \ifx\@let@token\bgroup.\else
  \ifx\@let@token\egroup.\else
  \ifx\@let@token\/.\else
  \ifx\@let@token\ .\else
  \ifx\@let@token~.\else
  \ifx\@let@token!.\else
  \ifx\@let@token,.\else
  \ifx\@let@token:.\else
  \ifx\@let@token;.\else
  \ifx\@let@token?.\else
  \ifx\@let@token/.\else
  \ifx\@let@token'.\else
  \ifx\@let@token).\else
  \ifx\@let@token-.\else
  \ifx\@let@token\@xobeysp.\else
  \ifx\@let@token\space.\else
  \ifx\@let@token\@sptoken.\else
   .\space
   \fi\fi\fi\fi\fi\fi\fi\fi\fi\fi\fi\fi\fi\fi\fi\fi\fi\fi}
\catcode`\@=12 % @ signs are no longer letters
\newcommand{\CL}[1]{$#1\%$~C.L\xdotspace}
\newcommand{\stru}[2]{%
   \relax\ifmmode\hbox{\vrule height#1 depth#2 width0pt}%
   \else\vrule height#1 depth#2 width0pt\fi}
\newcommand{\Ronum}[1]{\uppercase\expandafter{\romannumeral#1}}
\newcommand{\ronum}[1]{\expandafter{\romannumeral#1}}
\DeclareRobustCommand{\LaTeXZ}{%
  \LaTeX\kern-.05em4\kern-.1em
  {\raisebox{-0.2ex}{$\scriptstyle\text{ZEUS}$}}\xspace}
\newcommand{\fig}[1]{Fig.~\ref{fig-#1}}
\newcommand{\Fig}[1]{Figure~\ref{fig-#1}}
\newcommand{\figand}[2]{Figs.~\ref{fig-#1} and~\ref{fig-#2}}

\newcommand{\taband}[2]{Tables~\ref{tab-#1} and~\ref{tab-#2}}

\newcommand{\Sect}[1]{Section~\ref{sec-#1}}

\DeclareMathAlphabet{\mathbf}{OT1}{cmr}{bx}{sl}
\newcommand{\eVdist}{\kern-0.06667em}

\newcommand{\gev}{{\,\text{Ge}\eVdist\text{V\/}}}
\newcommand{\tev}{{\,\text{Te}\eVdist\text{V\/}}}
\newcommand{\pbi}{\,\text{pb}^{-1}}

\newcommand{\met}{\,\text{m}}

\newcommand{\mm}{\,\text{mm}}
\newcommand{\cm}{\,\text{cm}}

\newcommand{\Tesla}{\,\text{T}}

\newcommand{\slashfrac}[2]{%
  \raisebox{0.5ex}{\ensuremath #1}\kern-0.12em/\kern-0.08em
  \raisebox{-.8ex}{\ensuremath #2}}
\newcommand{\sqr}[3]{%
    {\vcenter{\hrule height.#3ex\hbox{\vrule width.#2ex height#1ex
     \kern#1ex\vrule width.#3ex}\hrule height.#2ex}}}

\catcode`\@=11 % @ signs are now treated as letters
\newcommand{\parenbar}{\mathpalette\p@renb@r}
\def\p@renb@r#1#2{\vbox{%
  \ifx#1\scriptscriptstyle \dimen@.7em\dimen@ii.2em\else
  \ifx#1\scriptstyle \dimen@.8em\dimen@ii.25em\else
  \dimen@1em\dimen@ii.4em\fi\fi \offinterlineskip
  \ialign{\hfill##\hfill\cr
    \vbox{\hrule width\dimen@ii}\cr
    \noalign{\vskip-.3ex}%
    \hbox to\dimen@{$\mathchar300\hfil\mathchar301$}\cr
    \noalign{\vskip-.3ex}%
    $#1#2$\cr}}}
\catcode`\@=12 % @ signs are no longer letters

\newcommand{\rnge}{\hbox{$\,\text{--}\,$}}

\newcommand{\IP}{{\rm I$\kern-0.01667em$P}\xspace}

\newcommand{\LQ}{{\rm LQ}}

\mathchardef\qsm=63
\mathchardef\pls=43
\mathchardef\mns=512
\mathchardef\plm=518
\mathchardef\eql=61
\mathchardef\smallleft=300
\mathchardef\smallright=301
\mathchardef\les=316
\mathchardef\gre=318
\mathchardef\leq=532
\mathchardef\grq=533
\catcode`\@=11 % @ signs are now treated as letters
\newcounter{pict@width}
\newcounter{pict@height}
\newlength{\pict@scale}
\newcommand{\psfigadd}[4]{%
\setcounter{pict@width}{1*\ratio{#2+\pict@scale/2}{\pict@scale}}
\setcounter{pict@height}{1*\ratio{#3+\pict@scale/2}{\pict@scale}}
\setlength{\unitlength}{\pict@scale}
\hbox to #2{\hspace{-\fill}\begin{picture}(\thepict@width,\thepict@height)
\put(0,0){\psfig{figure=#1,width=#2,height=#3,clip=}}
\SetScale{0.283466457}
\SetWidth{1.763889}
{#4}
\end{picture}}
}
\newcounter{pict@widthfst}
\newcounter{pict@widthscd}
\newcounter{pict@widthtot}
\newcommand{\psfigaddtwo}[7]{%
\setcounter{pict@widthfst}{1*\ratio{#2+\pict@scale/2}{\pict@scale}}
\setcounter{pict@widthscd}{1*\ratio{#2+#4+\pict@scale/2}{\pict@scale}}
\setcounter{pict@widthtot}{1*\ratio{#2+#4+#6+\pict@scale/2}{\pict@scale}}
\setcounter{pict@height}{1*\ratio{#3+\pict@scale/2}{\pict@scale}}
\setlength{\unitlength}{\pict@scale}
\hbox{\hspace{-\fill}\begin{picture}(\thepict@widthtot,\thepict@height)
\put(0,0){\psfig{figure=#1,width=#2,height=#3,clip=}}
\put(\thepict@widthscd,0){\psfig{figure=#5,width=#6,height=#3,clip=}}
\SetScale{0.283466457}
\SetWidth{1.763889}
{#7}
\end{picture}}
}
\newcommand{\psfigror}[4]{%
\setcounter{pict@width}{1*\ratio{#2+\pict@scale/2}{\pict@scale}}
\setcounter{pict@height}{1*\ratio{#3+\pict@scale/2}{\pict@scale}}
\setlength{\unitlength}{\pict@scale}
\hbox{\begin{picture}(\thepict@width,\thepict@height)
\put(0,\thepict@height){\psfig{figure=#1,width=#3,height=#2,clip=,angle=270}}
\SetScale{0.283466457}
\SetWidth{1.763889}
{#4}
\end{picture}}
}
\newcommand{\psfigrol}[4]{%
\setcounter{pict@width}{1*\ratio{#2+\pict@scale/2}{\pict@scale}}
\setcounter{pict@height}{1*\ratio{#3+\pict@scale/2}{\pict@scale}}
\setlength{\unitlength}{\pict@scale}
\hbox{\begin{picture}(\thepict@width,\thepict@height)
\put(0,0){\psfig{figure=#1,width=#3,height=#2,clip=,angle=90}}
\SetScale{0.283466457}
\SetWidth{1.763889}
{#4}
\end{picture}}
}
\catcode`\@=12 % @ signs are no longer letters
\newlength\listtextwidth

\catcode`\@=11 % @ signs are now treated as letters
\newlength{\@tabfninsert}
\newlength{\@tabfnwidth}
\newcommand{\tabfootnote}[2]{%
  \setlength{\@tabfninsert}{0.8em}
  \setlength{\@tabfnwidth}{\textwidth}
  \addtolength{\@tabfnwidth}{-\@tabfninsert}
  \addtolength{\@tabfnwidth}{-0.4em}
  \noindent\makebox[\@tabfninsert][r]{\footnotesize$^{#1}$\hfil}\hfill%
  \parbox[t]{\@tabfnwidth}{\footnotesize #2\hfill}}
\catcode`\@=12 % @ signs are no longer letters

\newcommand{\br}[1]{{\beta_{#1}}}
\newcommand{\MLQ}{M_{\LQ}}

\newcommand {\ptmiss}{\mbox{$\not\hspace{-0.55ex}{P}_t$}}

\newcommand{\rp}{$\not$\kern-0.pt$R_p$}

\begin{document}

\prepnum{{DESY--05--016}}

\title{
Search for lepton-flavor violation at HERA
}

\author{ZEUS Collaboration}
\date{January 2005}

\abstract{
A search for lepton-flavor-violating interactions  $e p \to \mu X $ and $e p\to \tau X $ has been performed with the ZEUS detector using the entire HERA~I data sample, corresponding to an integrated luminosity of $130\pbi$. The data were taken at center-of-mass energies, $\sqrt{s}$, of $300$ and $318\gev$. No evidence of lepton-flavor violation was found, and constraints were derived on leptoquarks (LQs) that could mediate such interactions. For LQ masses below $\sqrt{s}$, limits were set on $\lambda_{eq_1} \sqrt{\beta_{\ell q}}$, where $\lambda_{eq_1}$ is the coupling of the LQ to an electron and a first-generation quark $q_1$, and $\beta_{\ell q}$ is the branching ratio of the LQ to the final-state lepton $\ell$ ($\mu$ or $\tau$) and a quark $q$. For LQ masses much larger than  $\sqrt{s}$, limits were set on  the four-fermion interaction term $\lambda_{e q_\alpha} \lambda_{\ell q_\beta} / M_{\mathrm{LQ}}^2$ for LQs that couple to an electron and a quark $q_\alpha$ and to a lepton $\ell$ and a quark $q_\beta$, where $\alpha$ and $\beta$ are quark generation indices.
Some of the limits are also applicable to lepton-flavor-violating processes mediated by squarks in $R$-Parity-violating supersymmetric models.
In some cases, especially when a higher-generation quark is involved and for the process $e p\to \tau X $, the ZEUS limits are the most stringent to date.
}

\makezeustitle

\def\3{\ss}

\pagenumbering{Roman}

\begin{center}
{                      \Large  The ZEUS Collaboration              }
\end{center}
\pagenumbering{Roman}                                                                              
                                    % this "%"s are for cosmetics only                             
%\begin{document}                                                                                   
                                                   %                                               
%\begin{center}                                                                                     
%{                      \Large  The ZEUS Collaboration              }                               
%\end{center}                                                                                       
                                                   %                                               
  S.~Chekanov,                                                                                     
  M.~Derrick,                                                                                      
  S.~Magill,                                                                                       
  S.~Miglioranzi$^{   1}$,                                                                         
  B.~Musgrave,                                                                                     
  \mbox{J.~Repond},                                                                                
  R.~Yoshida\\                                                                                     
 {\it Argonne National Laboratory, Argonne, Illinois 60439-4815}, USA~$^{n}$                       
\par \filbreak                                                                                     
  M.C.K.~Mattingly \\                                                                              
 {\it Andrews University, Berrien Springs, Michigan 49104-0380}, USA                               
\par \filbreak                                                                                     
  N.~Pavel, A.G.~Yag\"ues Molina \\                                                                
  {\it Institut f\"ur Physik der Humboldt-Universit\"at zu Berlin,                                 
           Berlin, Germany}
% ...
\newpage                                                                                           
$^{\    1}$ also affiliated with University College London, UK \\                                  
% ...
                                                           %                                       
                                                           %                                       
% \par         % if index listing & table fit to 1 page, put gap here                              
\newpage   % alternatively: go to newpage, if page is too small                                    
                                                           %                                       
% \institute_references_start    % do not touch or move this line !                                
                                                           %                                       
\begin{tabular}[h]{rp{14cm}}                                                                       
$^{a}$ &  supported by the Natural Sciences and Engineering Research Council of Canada (NSERC) \\  
$^{b}$ &  supported by the German Federal Ministry for Education and Research (BMBF), under        
          contract numbers HZ1GUA 2, HZ1GUB 0, HZ1PDA 5, HZ1VFA 5\\                                
$^{c}$ &  supported in part by the MINERVA Gesellschaft f\"ur Forschung GmbH, the Israel Science   
          Foundation (grant no. 293/02-11.2), the U.S.-Israel Binational Science Foundation and    
          the Benozyio Center for High Energy Physics\\                                            
$^{d}$ &  supported by the German-Israeli Foundation and the Israel Science Foundation\\           
$^{e}$ &  supported by the Italian National Institute for Nuclear Physics (INFN) \\                
$^{f}$ &  supported by the Japanese Ministry of Education, Culture, Sports, Science and Technology 
          (MEXT) and its grants for Scientific Research\\                                          
$^{g}$ &  supported by the Korean Ministry of Education and Korea Science and Engineering          
          Foundation\\                                                                             
$^{h}$ &  supported by the Netherlands Foundation for Research on Matter (FOM)\\                   
$^{i}$ &  supported by the Polish State Committee for Scientific Research, grant no.               
          620/E-77/SPB/DESY/P-03/DZ 117/2003-2005 and grant no. 1P03B07427/2004-2006\\             
$^{j}$ &  partially supported by the German Federal Ministry for Education and Research (BMBF)\\   
$^{k}$ &  supported by RF Presidential grant N 1685.2003.2 for the leading scientific schools and  
          by the Russian Ministry of Education and Science through its grant for Scientific        
          Research on High Energy Physics\\                                                        
$^{l}$ &  supported by the Spanish Ministry of Education and Science through funds provided by     
          CICYT\\                                                                                  
$^{m}$ &  supported by the Particle Physics and Astronomy Research Council, UK\\                   
$^{n}$ &  supported by the US Department of Energy\\                                               
$^{o}$ &  supported by the US National Science Foundation\\                                        
$^{p}$ &  supported by the Polish Ministry of Scientific Research and Information Technology,      
          grant no. 112/E-356/SPUB/DESY/P-03/DZ 116/2003-2005 and 1 P03B 065 27\\                  
$^{q}$ &  supported by FNRS and its associated funds (IISN and FRIA) and by an Inter-University    
          Attraction Poles Programme subsidised by the Belgian Federal Science Policy Office\\     
\end{tabular}                                                                                      
                                                           %                                       
% \institute_references_end     % do not touch or move this line !                                 
                                                           %                                       
%\end{document}                                                                                     

\newpage
\pagenumbering{arabic} 
\pagestyle{plain}

\section{Introduction}
\label{sec-int}
The recent observations of neutrino oscillations~\cite{prl:81:1562,prl:87:071301} have shown that lepton-flavor violation (LFV) does occur in the neutrino sector. The LFV induced in the charged-lepton sector due to neutrino oscillations cannot be measured at existing colliders due to the low expected rate~\cite{app:b34:5413}. However, there are many extensions of the Standard Model (SM) such as grand unified theories (GUT)~\cite{pr:d10:275,*prl:32:438,*prep:72:185}, supersymmetry (SUSY)~\cite{prep:110:1,*prep:117:75}, compositeness~\cite{pl:b153:101,*pl:b167:337} and technicolor~\cite{np:b155:237,*np:b168:69,*pr:d20:3404,*prep:74:277} that predict possible $e\to\mu$ or $e\to\tau$ transitions at detectable rates.

In many theories, LFV occurs only in the presence of a particular quark generation. At the HERA $ep$ collider, lepton-flavor-violating interactions can be observed in the reaction $ep\to \ell X$, where $\ell$ is a $\mu$ or $\tau$. The presence of such processes, which can be detected almost without background, would clearly be a signal of physics beyond the Standard Model. This search is sensitive to all quark generations for LFV occurring between $e$ and $\mu$ or $\tau$. Strong constraints on LFV also arise from measurements of rare lepton and meson decay, muon-electron conversion on nuclei, etc.~\cite{zfp:c61:613,*pr:d62:055009,*Herz:2002gq,*pr:d66:010001,*Aubert:2003pc,*Yusa:2004gm,*prl:93:241802}; nevertheless, HERA generally has a competitive sensitivity, and better sensitivity in the case of $e\rnge\tau$ transition when a second- or third-generation quark is involved.

\begin{figure}[p]
\begin{center}
\epsfig{figure=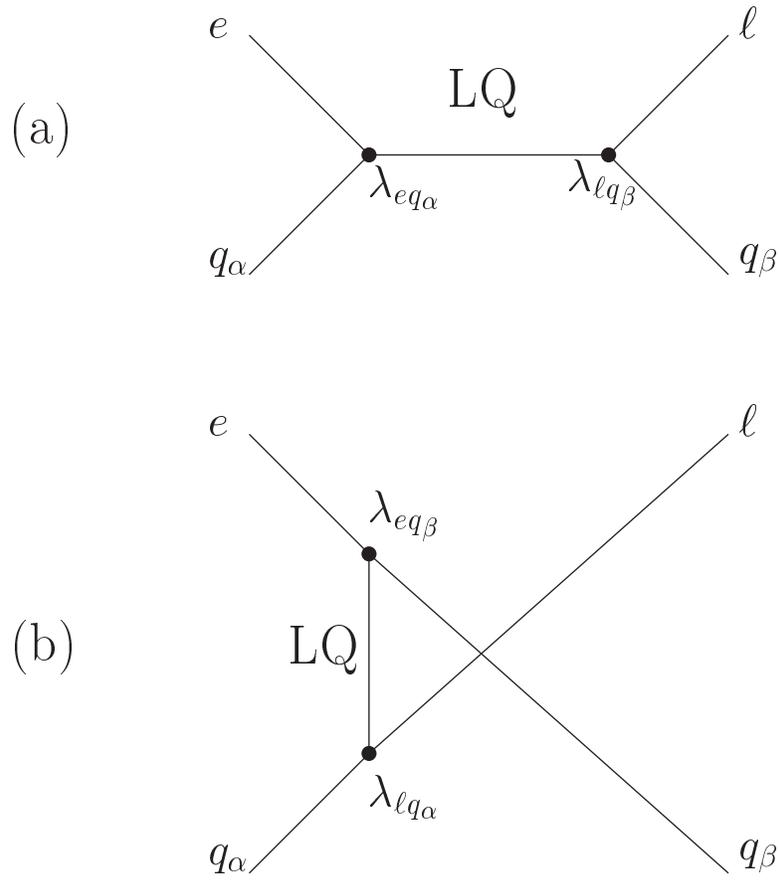,height=12cm }
\end{center}
\caption{(a) $s$-channel and (b) $u$-channel diagrams contributing to LFV processes.
The subscripts $\alpha$ and $\beta$ denote the quark generations, and $\ell$ is either a $\mu$ or a $\tau$.} 
\label{fig-LQFEY}
\end{figure}

In this search, no evidence for LFV was found. The Buchm\"uller-R\"uckl-Wyler (BRW) leptoquark (LQ) model~\cite{pl:b191:442} and supersymmetry with $R$-Parity violation are used to set limits from the search. Leptoquarks are bosons that carry both leptonic ($L$) and baryonic ($B$) numbers and have lepton-quark Yukawa couplings. Their fermionic number ($F=3B+L$) can be $F=0$ or $|F|=2$. Such bosons arise naturally in unified theories that arrange quarks and leptons in common multiplets. A LQ that couples both to electrons and to higher-generation leptons would induce LFV in $ep$ collisions through the $s$- and $u$-channel processes shown in \fig{LQFEY}. The same processes can also be mediated by squarks, the supersymmetric partners of quarks, in SUSY theories that violate $R$-Parity~\cite{pr:d40:2987}. A detailed description of the considered phenomenological scenarios and of the cross section assumptions used in this paper is given in a previous publication~\cite{pr:d65:92004}. 

Searches for LFV have been previously made at HERA~\cite{epj:c11:447,pr:d65:92004}. This analysis is based on the entire HERA I sample collected by ZEUS in the years $1994\rnge2000$, corresponding to an integrated luminosity of $130 \pbi$. These results supersede previous results published by ZEUS~\cite{zfp:c73:613,pr:d65:92004}, based on a sub-sample of the present data.

\section{The experimental conditions}
\label{sec-exp}
A detailed description of the ZEUS detector can be found elsewhere \cite{zeus:1993:bluebook}. In this section a brief outline of the main components used in this analysis is given: the central tracking detector (CTD)~\cite{nim:a279:290,*npps:b32:181,*nim:a338:254}, the uranium-scintillator calorimeter (CAL)~\cite{nim:a309:77,*nim:a309:101,*nim:a321:356,*nim:a336:23} and the forward muon detector (FMUON)~\cite{zeus:1993:bluebook}.

The CTD, which is immersed in a magnetic field of $1.43\Tesla$ provided by a superconducting solenoid, consists of 72~cylindrical drift chamber layers, organized in 9~superlayers covering the polar-angle\ZcoosysfnA ~region \mbox{$15^\circ<\theta<164^\circ$}. The transverse-momentum resolution for full-length tracks is $\sigma(p_T)/p_T=0.0058p_T\oplus0.0065\oplus0.0014/p_T$, with $p_T$
in $\gev$. The CTD was used to reconstruct tracks of isolated muons and charged $\tau$-decay products. It was also used to reconstruct the interaction vertex with a typical resolution of $4\mm$ ($1\mm$) in the $Z$ ($X$ and $Y$) coordinate.

The high-resolution uranium--scintillator calorimeter consists of three parts: the forward (FCAL), the barrel (BCAL) and the rear (RCAL) calorimeters. Each part is subdivided transversely into towers and longitudinally into one electromagnetic section (EMC) and either one (in RCAL) or two (in BCAL and FCAL) hadronic sections (HAC). The smallest subdivision of the calorimeter is called a cell.  The CAL energy resolutions, as measured under test-beam conditions, are $\sigma(E)/E=0.18/\sqrt{E}$ for electrons and $\sigma(E)/E=0.35/\sqrt{E}$ for hadrons ($E$ in $\gev$).

The FMUON detector, located between $Z=5\met$ and $Z=10\met$, consists of 6 planes of streamer tubes and 4 planes of drift chambers. The magnetic field of $1.6\Tesla$ produced by two iron toroids placed at about $9\met$ from the interaction point and the magnetic field of the iron yoke ($1.4\Tesla$) placed around the CAL enable the muon-momentum measurements to be made. The use of FMUON extends the acceptance for high-momentum muon tracks in the polar-angle region $8^\circ<\theta<20^\circ$.

The luminosity was measured using the process $ep\to e\gamma p$. The small-angle photons were measured by the luminosity detector~\cite{desy-92-066,*zfp:c63:391,*acpp:b32:2025}, a lead-scintillator calorimeter placed in the HERA tunnel at $Z=-107\met$.

\subsection{Kinematic quantities}
\label{sec-kin}
The total four-momentum in the CAL $(E,P_X,P_Y,P_Z)$ is defined as:
\[(\sum_i E_i,\sum_i E_i\sin\theta_i\cos\phi_i,\sum_iE_i\sin\theta_i \sin\phi_i, \sum_i E_i\cos\theta_i),\]
where $E_i$ is the energy measured in the $i^{\mathrm{th}}$ calorimeter cell. The angular coordinates $\theta_i$ and $\phi_i$ of the $i^{\mathrm{th}}$ cell are measured with respect to the reconstructed event vertex. The absolute value of the missing transverse momentum, $\ptmiss$, is given by $\sqrt{P_X^2+P_Y^2}$, while the transverse energy, $E_t$, is defined as $\sum_i E_i \sin\theta_i$.

Another relevant quantity used in this analysis is $E-P_Z=\sum_i E_i(1-\cos\theta_i)$. In the initial state, $E-P_Z=2E_e$, where $E_e$ is the electron beam energy of $27.5\gev$. If only particles in the very forward direction (proton beam), which give negligible contribution to this variable, are lost, as in NC DIS events, $E-P_Z\sim55\gev$ is measured in the final state.  

Jets are reconstructed using the $k_T$ cluster algorithm~\cite{np:b406:187} in the inclusive mode~\cite{pr:d48:3160}; only jets with transverse momentum greater than $4\gev$ are considered.

\section{Data samples and Monte Carlo simulation}
\label{sec-datamc}
The data used in this analysis were collected in the years 1994--2000. The total integrated luminosity was $112.8\pm 2.2\pbi$ with $e^+p$ collisions at the center-of-mass energy of $300$ and $318\gev$ and $16.7\pm 0.3\pbi$ with  $e^-p$ collisions at $318\gev$.

In the absence of a signal, limits were placed on LFV coupling strengths. The search is sensitive to any process with a final-state topology where the scattered electron of the $ep$ neutral current (NC) deep inelastic scattering (DIS) is replaced with a $\mu$ or a $\tau$. However, for the purpose of limit setting, the signal was  taken to be the LFV processes mediated by scalar or vector LQs of any mass.  These were simulated by the Monte Carlo (MC) generator \textsc{Lqgenep}~1.0~\cite{cpc:141:83}, which is based on the BRW model. The simulation of the hadronization and particle decays was performed using \textsc{Pythia}~6.1~\cite{cpc:82:74}.

Various MC samples were used to study the Standard Model background. Charged current (CC) and NC DIS events were simulated using \textsc{Djangoh}~1.1~\cite{spi:www:djangoh11}, an interface to the program \textsc{Heracles}~4.6.1~\cite{cpc:69:155:bis,*spi:www:heracles} and \textsc{Lepto}~6.5.1~\cite{cpc:101:108}; \textsc{Herwig}~6.1~\cite{cpc:67:465} was used for photoproduction background simulation while lepton pair production was simulated with \textsc{Grape}~1.1~\cite{cpc:136:126}.

\section{{\boldmath${e-\mu}$} transition}
\label{sec-muchan}
The characteristic of such events is an isolated  muon with high transverse momentum, which is balanced by that of a jet in the transverse plane. An apparent missing transverse momentum, measured by the calorimeter, due to the penetrating muon is used for event selection. Further requirements were applied, as described below, to identify charged particles as muons.  

\subsection{Muon identification}
\label{sec-mufind}
The muon identification comprises two different methods, in two different angular regions, for the final-state $\mu$ candidate. The first was used in the polar-angle range $15^\circ<\theta<164^\circ$ and required that the following conditions were satisfied:
\begin{itemize}
\item  a CTD track pointing to the vertex with transverse momentum above $5\gev$ matching a calorimeter deposit compatible with a minimum-ionizing particle;
\item $D_{\mathrm{trk}}>0.5$ and $D_{\mathrm{jet}}>1$ where $D_{\mathrm{trk}}$ ($D_{\mathrm{jet}}$) is the distance in the $\eta-\phi$ plane between the track associated with the candidate muon and the closest track (jet) to the candidate;
\item candidate muons in the polar-angle region $115^\circ<\theta<130^\circ$ were excluded to eliminate background from electrons that lose much of their energy in the dead material at the transition between BCAL and RCAL. 
\end{itemize}
The second method was used for very forward muons ($8^\circ<\theta<20^\circ$) and required a reconstructed track in the FMUON detector with hits in at least 5 detector planes.

\subsection{Preselection}
\label{sec-mupres}
The trigger used in this analysis was identical to that used in CC DIS measurement described in detail elsewhere \cite{epj:c12:411}. It was based on a cut on $\ptmiss$ with a threshold lower than that used in the offline analysis. After applying timing and other cuts to reject background due to non-$ep$ collisions (cosmics and beam-gas interactions), the following preselection requirements were imposed:
\begin{itemize}
\item a reconstructed vertex with $|Z_{\mathrm{vtx}}|<50\cm$;
\item $\ptmiss>15\gev$;
\item no electron\footnote{Throughout this paper, ``electron'' is used generically to refer to $e^+$ as well as $e^-$.}
candidate with energy larger than $10\gev$~\cite{epj:c11:427}; this cut was used to suppress NC DIS processes in a region of potentially high background and negligible anticipated signal;
\item an isolated-muon candidate in the direction of the $\ptmiss$ ($\Delta\phi<20^\circ$, where $\Delta\phi$ is the difference between the azimuthal angles of the candidate muon and of the $\ptmiss$ vector).
\end{itemize}
After the preselection, the sample contained 20 data events, while $25.9\pm1.1$ were expected from SM MC, mainly from QED di-muon processes ($ep\to \mu^+ \mu^- X$). 

\subsection{Final selection}
\label{sec-mufinal} 
\begin{figure}[p]
\vfill
\begin{center}
\epsfig{file=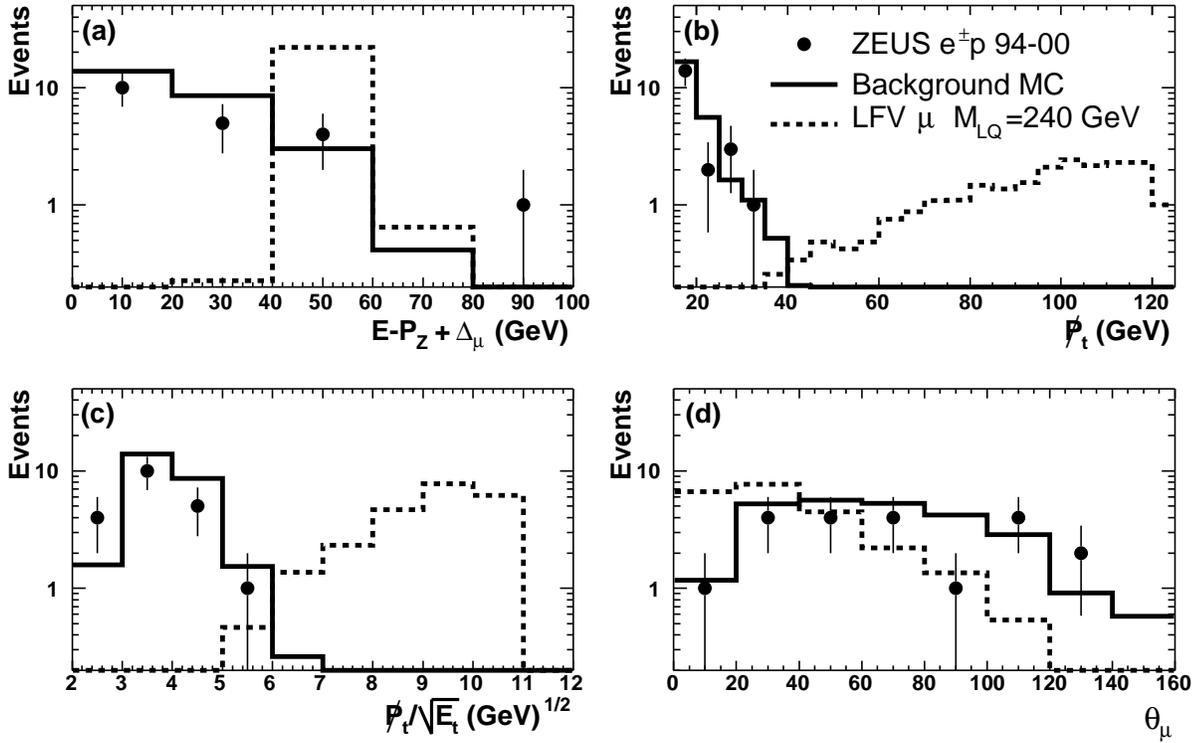, width=\textwidth}
\end{center}
\caption{Comparison between data (dots) and SM MC (solid line): (a) $E-P_{Z}+\Delta_{\mu}$, (b) $\ptmiss$, (c) $\ptmiss/\sqrt{E_t}$ and (d) polar angle of the muon, $\theta_{\mu}$, after the $\mu$-channel preselection. The dashed line represents the LFV signal due to a scalar LQ, with $M_{\mathrm{LQ}}=240\gev$, with an arbitrary normalization.
}
\label{fig-mupresel}
\vfill
\end{figure}

The cuts for the final selection were designed optimizing the sensitivity using signal and background simulations\cite{thesis:genta:05}. For this purpose a scalar LQ with a mass of $600\gev$, coupling to second-generation quarks, was taken as signal. Monte Carlo studies showed that this procedure results in good sensitivity for the whole range of LQ masses considered here. The following cuts were applied:
\begin{itemize}
\item $\ptmiss>20\gev$;
\item $\ptmiss/\sqrt{E_t}>3\sqrt{\gev}$; this cut was chosen to reject high-$E_t$ background events, where the small apparent $\ptmiss$ can arise from the finite energy-measurement resolution;
\item $E-P_{Z} +\Delta_\mu>45\gev$, where $\Delta_\mu=\ptmiss(1-\cos{\theta_\mu})/\sin{\theta_\mu}$, $\theta_\mu$ being the polar angle of the candidate muon; the quantity $\Delta_{\mu}$ represents the contribution to $E-P_Z$ carried by the muon, assuming that the transverse momentum of the muon is $\ptmiss$. 
\end{itemize}
\Fig{mupresel} shows the comparisons between data and MC expectations before the final selection. 
No event satisfied the final cuts, while $0.87\pm0.15$ were expected from the simulation of the SM background.

For LFV events mediated by resonant production of a leptoquark, the selection efficiency varied with the LQ mass, ranging from $39\%$ to $54\%$ for scalar LQs and from $47\%$ to $62\%$ for vector LQs with mass between $140$ and $300\gev$.

For leptoquarks with mass much greater than the center-of-mass energy the efficiency is typically lower than that for resonant LQs, because of the softer Bjorken-$x$ distributions of the initial-state quarks. In this case the efficiencies were almost independent of the LQ mass but depended on the generation of the initial-state quark. Sea quarks, with softer Bjorken-$x$ distribution  than valence quarks, result in a lower momentum of the final-state lepton, leading to a lower signal efficiency. Overall, the selection efficiency for high-mass LQs was in the range $20\rnge45\%$. 

\section{{\boldmath${e-\tau}$} transition}
\label{sec-tauchan}
Lepton-flavor-violating events leading to a final-state $\tau$ are characterized by a high-momentum isolated $\tau$ balanced by a jet in the transverse plane. Since the $\tau$ decays close to the interaction vertex, only its decay products are visible in  the detector. Due to the presence of at least one neutrino in all $\tau$-decay channels, a high value of $\ptmiss$ is expected. Therefore, for all the channels, the CC DIS trigger (as described in \Sect{mupres} for the muon channel) was used together with the following common preselection:
\begin{itemize} 
\item  $\ptmiss>15\gev$; 
\item a reconstructed vertex with $|Z_{\mathrm{vtx}}|<50\cm$.
\end{itemize}

\subsection{Leptonic $\tau$ decays}
\label{sec-taulepchan}
For $\tau$ leptons decaying into muons ($\tau\to\mu\nu_\mu\nu_\tau$), the same selection cuts as described in  \Sect{mufinal} were applied, since the event topology is very similar to that of LFV with $e\to\mu$ transitions.

For the  $\tau\to e \nu_e\nu_\tau$ channel, the final state is characterized by a high-energy isolated electron in the $\ptmiss$ direction; the following cuts were applied after the preselection:
\begin{itemize}
\item $20<E-P_Z<52\gev$;
\item total energy deposit in RCAL less than  $7\gev$;
\item $\ptmiss/\sqrt{E_t}>2.5\sqrt{\gev}$;
\item an electron with energy larger than $20\gev$ in the polar-angle region $8^\circ<\theta<125^\circ$ and in the $\ptmiss$ direction ($\Delta\phi<20^\circ$);
\item a jet with a transverse momentum above $25\gev$, back-to-back with respect to the electron ($\Delta\phi^{e-\mathrm{jet}}>160^\circ$) where $\Delta\phi^{e-\mathrm{jet}}$ is the difference between the azimuthal angles of the jet and of the electron.
\end{itemize}
No event was found in data, while $0.43\pm0.08$  were expected from SM MC.

\subsection{Hadronic $\tau$ decays}
\label{sec-tauhadchan}
The $\tau$ lepton, because of its small mass, typically decays with only one or three charged tracks with limited transverse spread. Since jets coming from hadronic $\tau$ decays must be separated  from a large background of QCD jets, a $\tau$ finder was employed to distinguish the $\tau$ jets from the quark- and gluon-induced jets. The algorithm exploits the fact that high-energy QCD jets usually have higher multiplicity and a larger internal transverse momentum than those for the decay products of the $\tau$.

\subsubsection{Tau identification}
A technique for $\tau$ identification~\cite{thesis:nguyen:02} was developed for a previous study~\cite{PL:B583:41} in which a small number of isolated-$\tau$ events were found in the data set identical to that used here. The longitudinally invariant $k_T$ cluster algorithm was used to identify jets. The jet shape was characterized by the following six observables~\cite{PL:B583:41}: the first ($R_{\mathrm{mean}}$) and the second ($R_{\mathrm{rms}}$) moment of the radial extension of the jet-energy deposition; the first ($L_{\mathrm{mean}}$) and the second ($L_{\mathrm{rms}}$) moment of the energy deposition in the direction along the jet axis; the number of subjets ($N_{\mathrm{subj}}$) within the jet resolved with a resolution criterion $y_{\mathrm{cut}}$ of $5\cdot10^{-4}$~\cite{jhep:09:009:bis,np:b421:545:bis}; the mass ($M_{\mathrm{jet}}$) of the jet calculated from the calorimeter cells associated with the jet. In order to separate the signal from the background, the six variables were combined into a discriminant $\mathcal{D}$, given, for any point in the phase space $\vec{x}(-\log(R_{\mathrm{mean}}),-\log(R_{\mathrm{rms}}),-\log(1-L_{\mathrm{mean}}),-\log(L_{\mathrm{rms}}),N_{\mathrm{subj}},M_{\mathrm{jet}})$, by:
\begin{equation*}
\mathcal{D}(\vec{x})=\frac{\rho_{\mathrm{sig}}(\vec{x})}{\rho_{\mathrm{sig}}(\vec{x})+\rho_{\mathrm{bkg}}(\vec{x})},
\end{equation*}
where $\rho_{\mathrm{sig}}$ and $\rho_{\mathrm{bkg}}$ are the density functions of the signal and the background, respectively. Such densities, sampled using MC simulations, were calculated using a method based on range searching~\cite{misc:hep-ph/00112224,*article:hep-ex/0211019}. Lepton-flavor-violating events in which the final-state $\tau$ decays into hadrons and a neutrino were used to simulate the signal. The background simulation was based on CC DIS MC events. For any given jet with phase space coordinates $\vec{x}$, the signal and the background densities were evaluated from the number of corresponding simulated signal and background jets in a 6-dimensional box of fixed size centered around $\vec{x}$. The $\tau$ signal tends to have a large discriminant value ($\mathcal{D}\to 1$) while the CC DIS background has a low discriminant value ($\mathcal{D}\to 0$).

\subsubsection{Preselection}
The following cuts were applied for the preselection of the hadronic $\tau$ decay channel:
\begin{itemize}
\item no electron candidate with energy larger than $10\gev$;
\item $E_t>45\gev$;
\item $15<E-P_{Z}<60\gev$;
\item total energy deposit in RCAL less than  $7\gev$;
\item a $\tau$-jet candidate as described below.
\end{itemize}
The $\tau$-jet candidate was required to have a transverse momentum greater than $15\gev$, to be within the CTD acceptance ($15^{\circ}<\theta<164^{\circ}$) and to have between one and three tracks pointing to the CAL energy deposit associated with the jet. Events with jets in the region between FCAL and BCAL ($36^\circ <\theta < 42^\circ$) were removed. In order to reject electrons from NC events, a cut of 0.95 was applied to the electromagnetic energy fraction of the jet ($f_{\mathrm{EMC}}$). In addition the jet was required to satisfy the condition $f_{\mathrm{LT}}+ f_{\mathrm{EMC}}<1.6$, where $f_{\mathrm{LT}}$ (the leading-track fraction) was defined as the ratio between the momentum of the most energetic track in the jet and the jet energy. The quantity $f_{\mathrm{LT}}+ f_{\mathrm{EMC}}$ is close to 2 for electrons, the main source of background that this cut is designed to reject.

\begin{figure}[p]
\vfill
\begin{center}
\epsfig{file=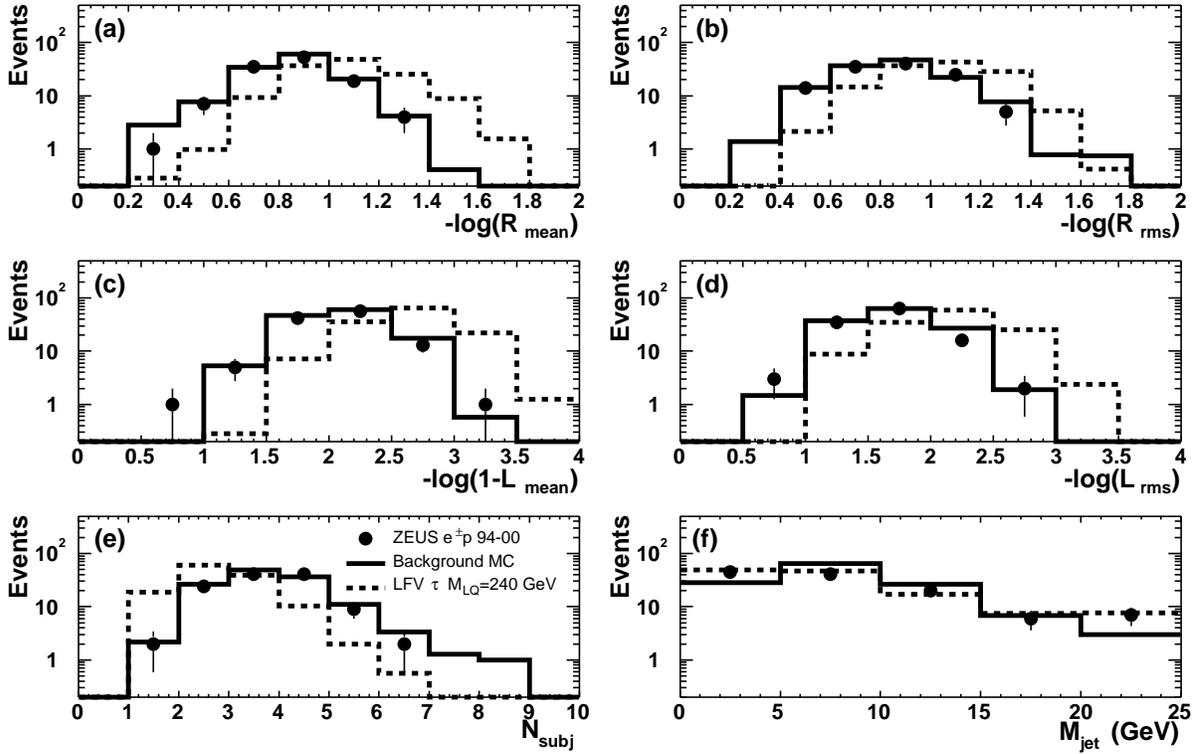, width=\textwidth}
\end{center}
\caption{Comparison between data (dots) and SM MC (solid line) for the variables used in the $\tau$ discriminant: (a) $-\log(R_{\mathrm{mean}})$; (b) $-\log(R_{\mathrm{rms}})$; (c) $-\log(1-L_{\mathrm{mean}})$; (d) $-\log(L_{\mathrm{rms}})$; (e) number of subjets, $N_{\mathrm{subj}}$; (f) jet mass, $M_{\mathrm{jet}}$, after the $\tau$-channel preselection (hadronic $\tau$ decays). The dashed line represents the LFV signal with arbitrary normalization.
}
\label{fig-taupresel-jetvar}
\vfill
\end{figure}

\begin{figure}[p]
\vfill
\begin{center}
\epsfig{file=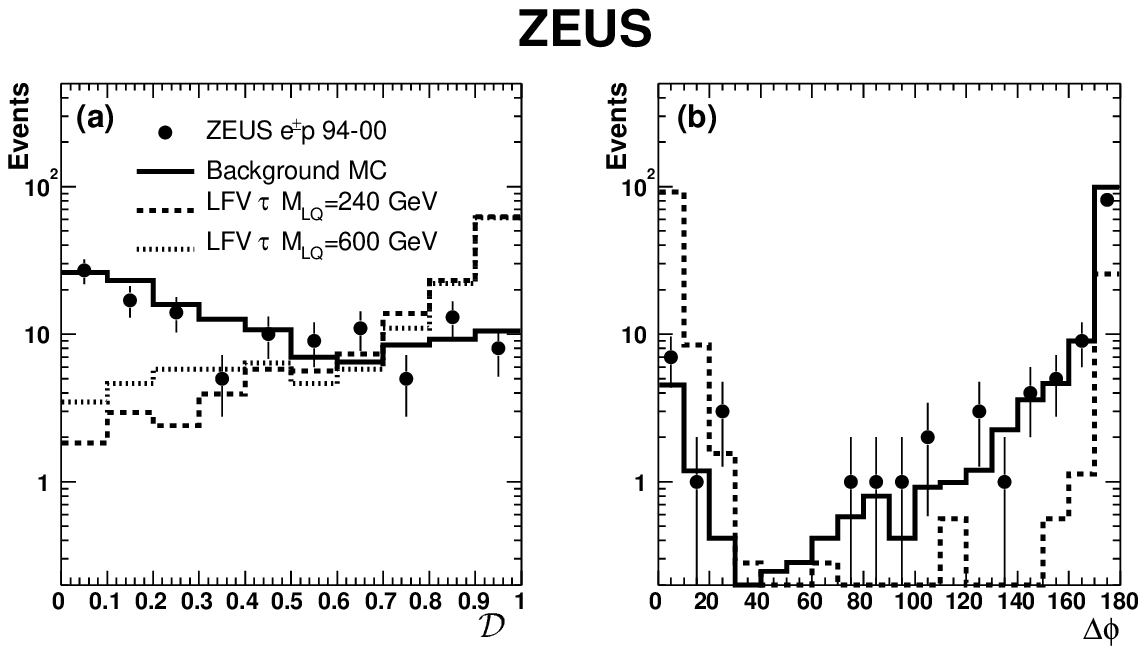, width=\textwidth}
\end{center}
\caption{Distribution of (a) the discriminant, $\mathcal{D}$, and (b) $\Delta\phi$, after hadronic $\tau$ decay preselection. The dots represent the data while the solid line is the SM prediction from MC. The LFV signal distribution for two different LQ masses, $240\gev$ (dashed line) and $600\gev$ (dash-dotted line), are also shown with arbitrary normalization. The distribution of $\Delta\phi$ for the $M_{\mathrm{LQ}}=600\gev$ LQ, which is similar to the $M_{\mathrm{LQ}}=240\gev$ LQ $\Delta\phi$ distribution, is omitted. The leptonic decay of the tau, or the tau jet outside the CTD acceptance, leads to events with $\Delta{\phi}>160^{\circ}$.
}
\label{fig-taupresel-d}
\vfill
\end{figure}

\Fig{taupresel-jetvar} shows the comparison, after the preselection, between data and MC for the jet discriminant variables. \Fig{taupresel-d} compares the discriminant and the $\Delta\phi$ distributions. Here, $\Delta\phi$ is the azimuthal angle between the candidate $\tau$-jet axis and the $\ptmiss$ vector. After the hadronic preselection, $119$ events were found in the data, while $131\pm 4$ were expected from SM processes,  mainly from CC DIS. The data distributions in \fig{taupresel-d} generally conform to those expected from SM backgrounds.

\subsubsection{Final selection}
\begin{figure}[p]
\vfill
\begin{center}
\epsfig{file=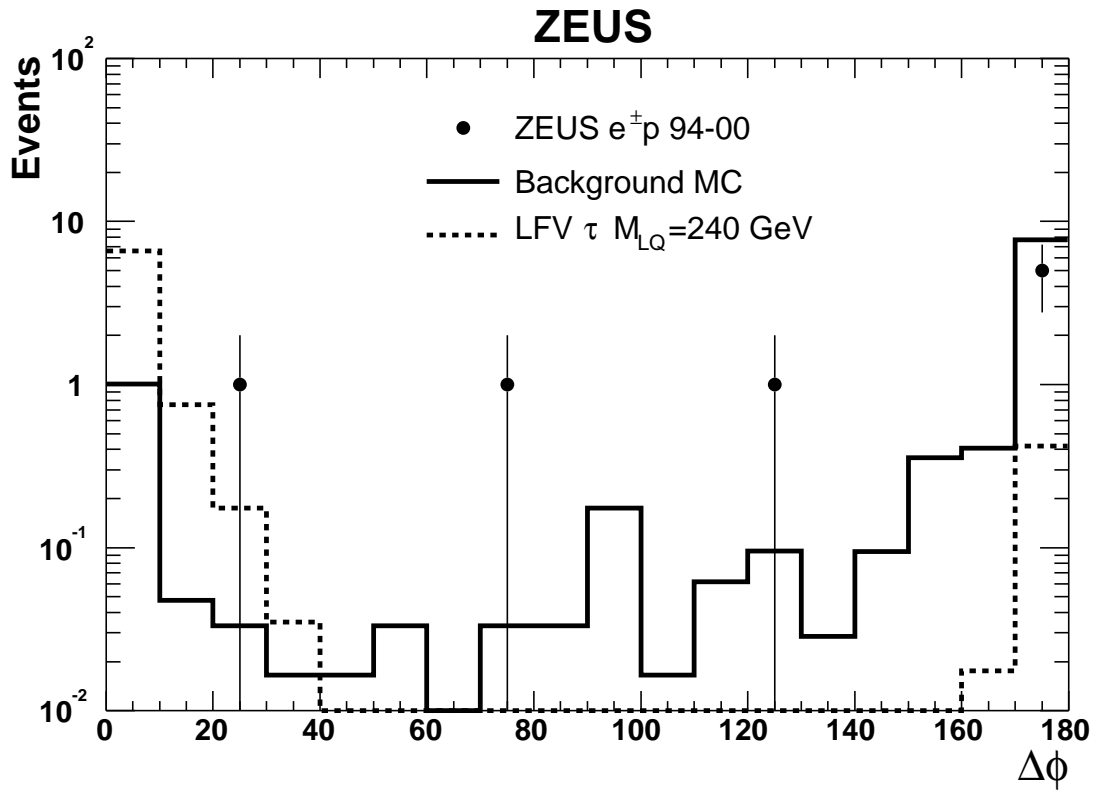, width=\textwidth}
\end{center}
\caption{$\Delta\phi$ distribution of the events with $\mathcal{D}>0.9$ after hadronic $\tau$ decay preselection. Dots represent data while the solid line is the SM prediction from MC. The dashed line represents the signal with arbitrary normalization. The small fraction of the signal ($\sim 5\%$) with $\Delta\phi>160^{\circ}$ is due to events that have the jet from the $\tau$ outside the CTD acceptance. The two events from data that have $\Delta\phi=72^{\circ}$ and  $\Delta\phi=126^{\circ}$ are the two events found in a previous ZEUS search for isolated $\tau$ lepton events~\protect\cite{PL:B583:41}.}
%}
\label{fig-deltaphi}
\vfill
\end{figure}

For the final selection, the following additional cuts were applied to the events in \fig{taupresel-d}:
\begin{itemize}
\item $\mathcal{D}>0.9$;
\item the  $\tau$-jet candidate was required to be aligned in azimuth with the direction of the $\ptmiss$ ($\Delta \phi<20^{\circ}$).
\end{itemize}
The discriminant cut was tuned to optimize the separation power, $S=\epsilon_{\mathrm{sig}} \cdot \sqrt{R}$ (where $\epsilon_{\mathrm{sig}}$ is the signal efficiency and $R=1/\epsilon_{\mathrm{bg}}$ is the background rejection), for a scalar LQ with a mass of $240\gev$~\cite{thesis:genta:05}. In \fig{deltaphi}, the $\Delta\phi$ distribution of the 8 events with $\mathcal{D}>0.9$ is shown compared to the SM expectation ($10.2\pm0.9$ events).

After imposing the final cut on $\Delta\phi$, no data events remained in the hadronic decay channel, while $1.1\pm0.5$ were expected from MC.

\subsection{Summary on $e \to \tau$ search}
No candidate was found in the data for any of the three $\tau$-decay channels, while $2.3\pm0.5$ were predicted by Standard Model simulations.

The combined selection efficiency for low-mass ($M_{\mathrm{LQ}}<\sqrt{s}$) scalar (vector) LQs was in the range of $22-29\%$ ($23-34\%$), while for high-mass ($M_{\mathrm{LQ}}\gg\sqrt{s}$)  LQs it was $4\rnge20\%$. As is the case for the $e\to\mu$ transition discussed above, the significant efficiency drop for high-mass LQs is due to the softer Bjorken-x distribution of the initial state quarks.

\section{Results}
Since no evidence of lepton-flavor-violating interactions was found, limits at \CL{95} were set -- using a Bayesian approach~\cite{proc:clw:2000:237:bis} that assumes a flat prior for the signal cross section -- on the processes $ep\to \mu X$ and $ep\to \tau X$ mediated by a leptoquark.

In the low-mass case, limits on the cross section were converted, using the narrow-width approximation, into limits on $\lambda_{eq_1}\sqrt{\beta_{\ell q}}$, where $\lambda_{eq_1}$ is the coupling between the leptoquark, the electron and a first-generation quark, while $\beta_{\ell q}$ is the branching ratio of the leptoquark into a lepton $\ell$ and a quark ($u$, $d$, $s$, $c$, $b$).
For high-mass leptoquarks, the cross-section limits were converted, using the contact-interaction approximation, into limits on $\lambda_{eq_\alpha}\lambda_{\ell q_\beta}/M^2_{\mathrm{LQ}}$, where $\alpha$ and  $\beta$ are quark generation indices. The cross sections were evaluated using the CTEQ5~\cite{pr:d55:1280} parton densities, taking into account the QED initial-state radiation, and, for low-mass scalar leptoquarks, NLO QCD corrections.

\subsection{Systematic uncertainties}
The following sources of systematic uncertainties are dominant:
\begin{itemize}
\item the calorimeter energy-scale uncertainty ($2\%$). The resulting variation in the signal efficiency for the muon ($\tau$) channel is less than  $1\%$ ($3\%$) for low-mass leptoquarks and less than $5\%$ for high-mass leptoquarks;   
\item the luminosity uncertainty: $1.5\%$ for the 1994-97 $e^+p$ data, $1.8\%$ for the 1998-99 $e^-p$ data and $2.2\%$ for the 1999-2000 $e^+p$ data;
\item Systematics related to the parton-density functions (PDF) have been calculated using the 40 eigenvector sets, provided by CTEQ~6.1~\cite{Pumplin:2002vw,*Stump:2003yu}, that characterize the PDF uncertanties. This contributes to the dominant uncertainty  for low-mass leptoquarks, especially when a $d$ quark is involved and the LQ mass approaches the HERA kinematic limit. 
The effect of this uncertainty on the LQ limits is given in more detail elsewhere~\cite{thesis:genta:05}.
\end{itemize}
The uncertainties related to muon and tau identification were evaluated following the methods described elsewhere~\cite{thesis:turcato:03,PL:B583:41} and were found to be small. The systematic uncertainties have been included in the limit calculation assuming a Gaussian distribution for their probability densities. For low-mass LQs, the effect of the inclusion of systematic uncertainties is the largest at the highest masses and the limit on the coupling increases by less than  $7\%$ at $250\gev$. The effect is very small for high-mass LQs (below 1\%).

\subsection{Low-mass leptoquark and squark limits}
\begin{figure}[p]
\vfill
\begin{center}
\epsfig{file=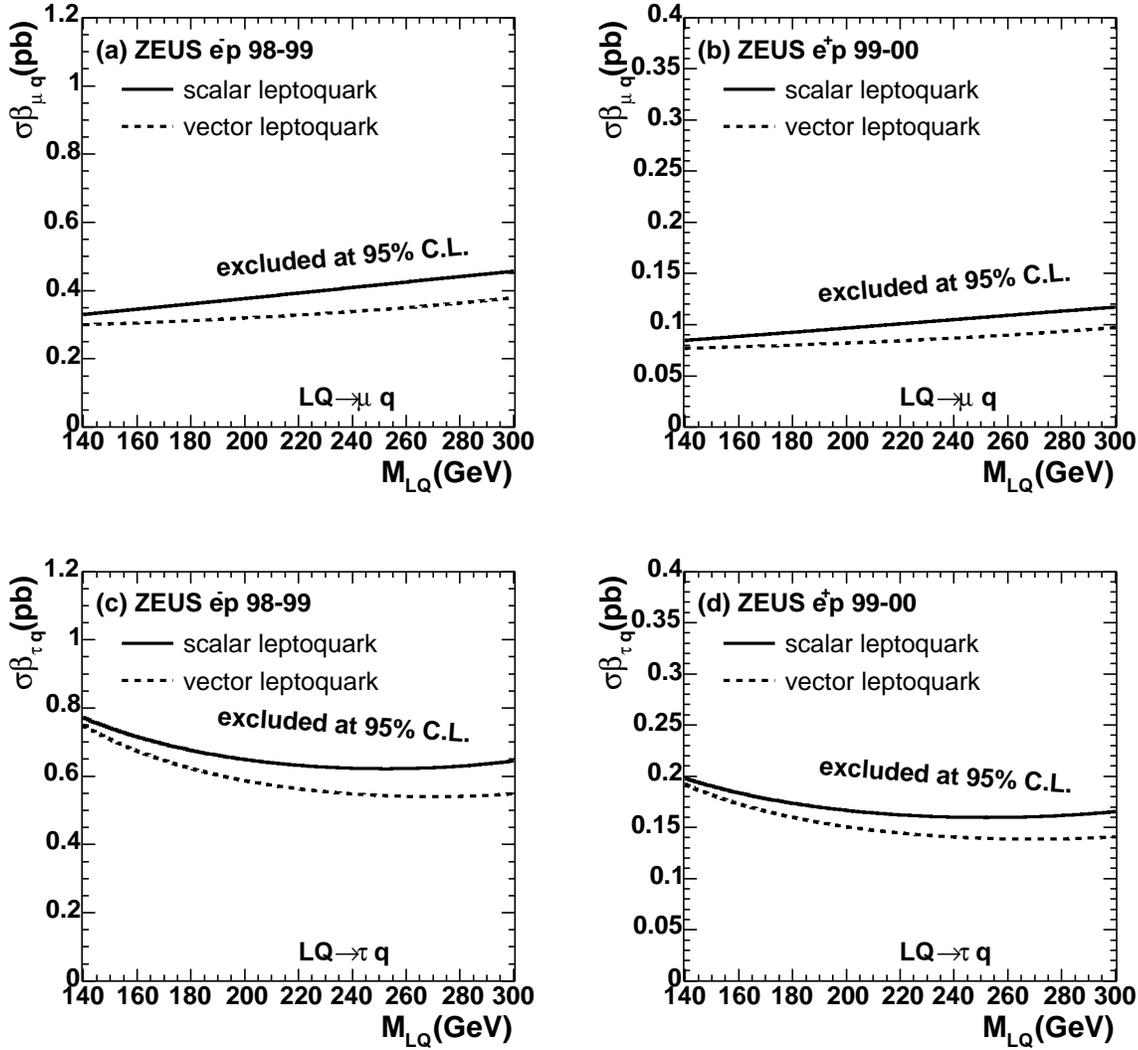, width=\textwidth}
\end{center}
\caption{The \CL{95} upper limits for $\sigma\beta_{\ell q}$ as a function of $M_{\mathrm{LQ}}$ for scalar(full line) and vector (dashed line) LQs: (a) $F=2$ $\mathrm{LQ}\to\mu q$; (b) $F=0$ $\mathrm{LQ}\to\mu q$; (c) $F=2$ $\mathrm{LQ}\to\tau q$; (d) $F=0$ $\mathrm{LQ}\to\tau q$. A subset of $e^+p$ data (99-00, corresponding to the higher center-of-mass energy, 318 GeV) has been used to obtain figures (b) and (d).
}
\label{fig-sigmabr}
\vfill
\end{figure}

\begin{figure}[p]
\vfill
\begin{center}
\epsfig{file=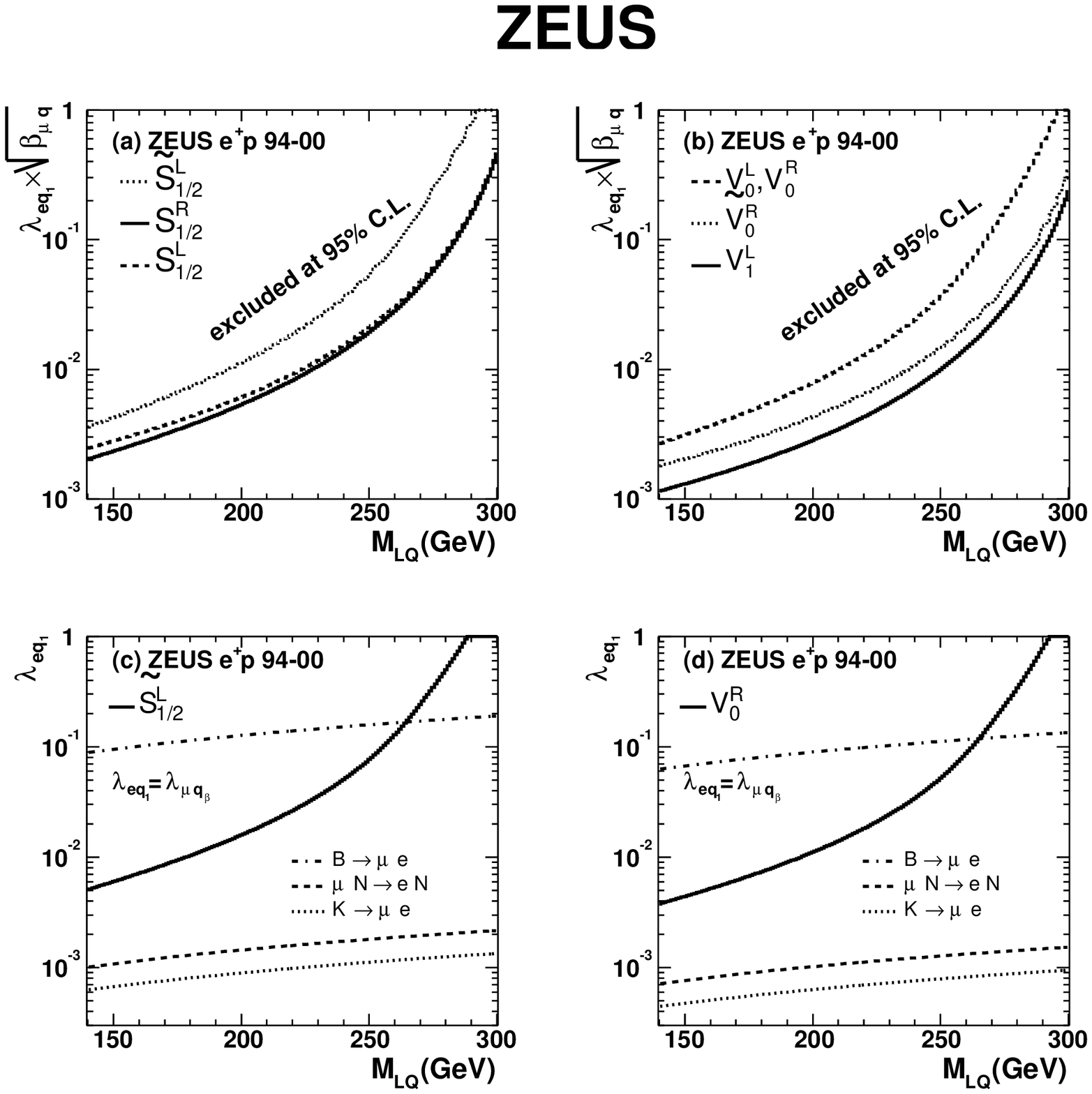, width=\textwidth}
\end{center}
\caption{Limits for $F=0$ low-mass LQs in the $\mu$ channel obtained from $e^{+}p$ collisions. The upper plots show \CL{95} limits on $\lambda_{eq_1}\times\sqrt{\beta_{\mu q}}$ for (a) scalar and (b) vector LQs. In the lower plots, ZEUS limits on $\lambda_{eq_1}$ for a representative (c) scalar and (d) vector LQ are compared to the indirect constraints from low-energy experiments~\protect\cite{zfp:c61:613,*pr:d62:055009,*pr:d66:010001,*Herz:2002gq,*Aubert:2003pc,*Yusa:2004gm,*prl:93:241802}, assuming $\lambda_{eq_1}=\lambda_{\mu q_{\beta}}$.
}
\label{fig-mu_f0}
\vfill
\end{figure}

\begin{figure}[p]
\vfill
\begin{center}
\epsfig{file=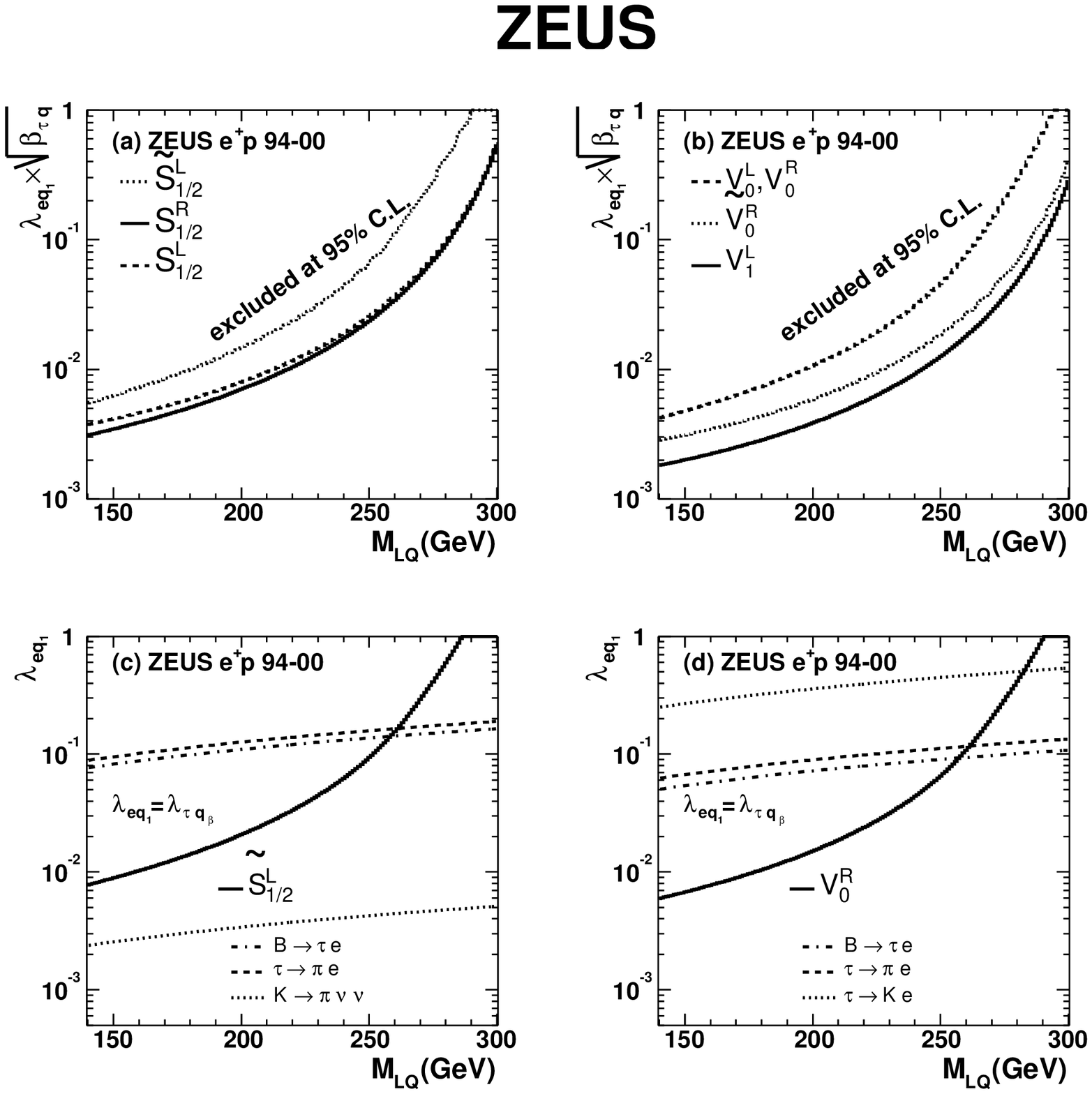, width=\textwidth}
\end{center}
\caption{Limits for $F=0$ low-mass LQs in the $\tau$ channel obtained from $e^{+}p$ collisions. The upper plots show \CL{95} limits on $\lambda_{eq_1}\times\sqrt{\beta_{\tau q}}$ for (a) scalar and (b) vector LQs. In the lower plots, ZEUS limits on $\lambda_{eq_1}$ for a representative (c) scalar and (d) vector LQ are compared to the indirect constraints  from low-energy experiments~\protect\cite{zfp:c61:613,*pr:d62:055009,*pr:d66:010001,*Herz:2002gq,*Aubert:2003pc,*Yusa:2004gm,*prl:93:241802}, assuming $\lambda_{eq_1}=\lambda_{\tau q_{\beta}}$.
}
\label{fig-tau_f0}
\vfill
\end{figure}

To illustrate the sensitivity of this search, \CL{95} upper limits on the cross section times the branching ratio, $\sigma\beta_{\ell q}$, for $F=0$ and $F=2$ leptoquarks are shown in \fig{sigmabr}; for the $e^+p$ case, only the subsample ($65\pbi$) with the higher $\sqrt{s}$ of $318\gev$ is used.
Upper limits on $\lambda_{eq_1}\sqrt{\beta_{\mu q}}$ are shown in \figand{mu_f0}{mu_f2} for $F=0$ and $|F|=2$ scalar and vector LQs, assuming resonantly produced leptoquarks as described by the BRW model. Since, for sufficiently large LQ masses, the cross section is dominated by electron valence-quark fusion, only $e^+p$ ($e^-p$) data were used to determine $F=0$ ($|F|=2$) LQ production limits. Similar considerations hold for the results shown for the $e-\tau$ channel in \figand{tau_f0}{tau_f2}. For couplings with electromagnetic strength ($\lambda_{eq_1}=\lambda_{\ell q_\beta}=0.3 \thickapprox \sqrt{4\pi\alpha}$), LQs with masses up to $299\gev$ are excluded (see \taband{LMF0}{LMF2}). Alternatively, for a fixed $M_{\mathrm{LQ}}$ of $250\gev$, values of $
\lambda_{eq_1}\sqrt{\beta_{\mu q}}$ and of  $\lambda_{eq_1}\sqrt{\beta_{\tau q}}$ down to $0.010$ and $0.013$, respectively, are excluded (see \taband{COUPF0}{COUPF2}).

Constraints on  $\lambda_{eq_1}\sqrt{\beta_{\ell q}}$ for $\tilde{S}_{1/2}^L$ and for $S_0^L$ can be interpreted as limits on $\lambda'_{1j1}\sqrt{\beta_{\tilde{u}^j\to\ell q}}$ and $\lambda'_{11k}\sqrt{\beta_{\tilde{d}^k\to\ell q}}$ for $\tilde{u}^j$ and $\tilde{d}^k$ $R$-Parity-violating squarks of generation $j$ and $k$, respectively \cite{np:b397:3}.

\subsection{High-mass leptoquark and squark limits}
\taband{HMF0MU}{HMF2MU} show the \CL{95} limits on $\lambda_{eq_\alpha}\lambda_{\mu q_\beta}/M^2_{\mathrm{LQ}}$ (third row of each cell) for $F=0$ and $|F|=2$ high-mass leptoquarks coupling to $eq_\alpha$ and $\mu q_\beta$. Limits were evaluated for all combinations of quark generations $\alpha$, $\beta$, except when a coupling to a $t$ quark is involved. \taband{HMF0TAU}{HMF2TAU} show the corresponding limits for LQs coupling to $eq_\alpha$ and $\tau q_\beta$.

Limits for $\tilde{S}_{1/2}^L$ LQs can also be interpreted as limits on $\lambda'_{1j\alpha}\lambda'_{ij\beta}/M^2_{\tilde{u}}$ for a $u$-type squark of generation $j$, where $i=2,3$ is the generation of the final-state lepton ($\mu$ or $\tau$). Similarly, limits for $S_0^L$ LQs can also be interpreted as limits on  $\lambda'_{1\alpha k}\lambda'_{i\beta k}/M^2_{\tilde{d}}$ for a $d$-type squark of generation $k$.

\section{Comparison with limits from other experiments}
\subsection{Low-energy experiments}
There are many constraints from low-energy experiments on lepton-flavor-violating processes coming from muon scattering and rare lepton or mesons decays~\cite{zfp:c61:613,*pr:d62:055009,*pr:d66:010001,*Herz:2002gq,*Aubert:2003pc,*Yusa:2004gm,*prl:93:241802}. Most can be converted into limits on  $\lambda_{eq_\alpha}\lambda_{\ell q_\beta}/M^2_{\mathrm{LQ}}$ for massive scalar or vector leptoquark exchange. In Tables~\ref{tab-HMF0MU}-\ref{tab-HMF2TAU}, the limits from such measurements are compared to the constraints from this analysis. For the $e-\mu$ transition, such indirect limits are very stringent and ZEUS limits are better only in a few cases involving  the $c$-quark. In the $e-\tau$ channel, ZEUS improves on the existing limits for many initial- and final-state quark combinations, especially when a quark of the second or third generation is involved. Assuming $\lambda_{eq_1}=\lambda_{\ell q_\beta}$, ZEUS limits on low-mass LQs can be compared to the limits from low-energy experiments. In \figand{mu_f0}{mu_f2}, limits on $\lambda_{eq_1}$ as a function of the LQ mass are compared to the limits from $e-\mu$ conversion in nuclei and from rare $K$- and $B$- meson decays. ZEUS limits are better or competitive with indirect limits up to $\sim250\gev$ when the quark in the final state is of the third generation. In \figand{tau_f0}{tau_f2}, the corresponding limits for the $\tau$ case are shown compared to constraints from rare $\tau$, $B$ or $K$ decays. ZEUS limits improve on low-energy results in most cases. 

\subsection{LFV and leptoquark searches at colliders}
Tevatron limits are complementary to those from HERA  since the cross sections at $p\bar{p}$ colliders do not depend on the Yukawa coupling, and  LQs are assumed to couple only with one lepton generation. Therefore, such experiments are sensitive to only a subset of the interactions considered here. The CDF and \DO collaborations exclude scalar LQs coupling exclusively to $\mu q$ with masses up to $202\gev$~\cite{prl:81:4806} and $200\gev$~\cite{prl:84:2088}, respectively. CDF performed an analysis searching for leptoquarks which couple exclusively to the third generation of leptons and excluded LQs with $M_{\mathrm{LQ}}<99\gev$ if $\beta_{\tau b}=1$. The \DO collaboration looking for $\nu\nu bb$ final states excluded LQs with masses below $94\gev$ if $\beta_{\nu b}=1$. The CDF collaboration also performed a search for a narrow resonance decaying to two charged leptons of different generation~\cite{unknown:2003jd}, observing no deviation from the SM expectation.

Searches for LFV interactions, not mediated by LQs, were performed by LEP experiments, looking for $e \mu$, $e \tau$ and $\mu \tau$ production in $e^+e^-$ annihilation at the $Z^0$ peak~\cite{pl:b254:293,*pr:216:253,*pl:b298:247,*pl:b316:427}; the OPAL collaboration extended the search to higher energy using LEP2 data~\cite{pl:b519:23}. Also in this case, no significant deviation from the SM expectation was found.
\section{Conclusions}
The data taken by the ZEUS experiment at HERA in $e^+p$ and $e^-p$ interactions at center-of-mass energies of $300\gev$ and $318\gev$ during the years 1994--2000 corresponding to an integrated luminosity of $130\pbi$ were analyzed for lepton-flavor violation. Searches in both $\mu$ and $\tau$ channels were performed. No evidence of lepton-flavor-violating interactions was found. For masses lower than the center-of-mass energy, limits at \CL{95} were set on $\lambda_{e q_1}\sqrt{\beta_{\ell q}}$ for leptoquark bosons as a function of the mass. For a coupling constant of electromagnetic strength ($\lambda_{e q_1}=\lambda_{\ell q_\beta}=0.3$), mass limits between $257$  and $299\gev$ were set, depending on the LQ type. For $M_{\mathrm{LQ}}=250\gev$, upper limits on  $\lambda_{e q_1}\sqrt{\beta_{\mu q}}$ ($\lambda_{e q_1}\sqrt{\beta_{\tau q}}$) in the range $0.010\rnge0.12$ ($0.013\rnge0.15$) were set.

For LQs with $M_{\mathrm{LQ}}\gg\sqrt{s}$, upper limits on $\lambda_{eq_\alpha}\lambda_{\mu q_\beta}/M^2_{\mathrm{LQ}}$ and $\lambda_{eq_\alpha}\lambda_{\tau q_\beta}/M^2_{\mathrm{LQ}}$  were calculated for all combinations of initial- and final-state quark generations.

Some of the limits also apply to $R$-Parity-violating squarks.
In many cases, especially in the $\tau$-channel, ZEUS limits are more stringent than any other limit published to date.

\section{Acknowledgments}
We would like to thank the DESY Directorate for their strong support and encouragement. The remarkable achievements of the HERA machine group were essential for the successful completion of this work and are greatly appreciated. The design, construction and installation of the ZEUS detector have been made possible by the effort of many people who are not listed as authors.
%\vfill\eject

\providecommand{\etal}{et al.\xspace}
\providecommand{\coll}{Coll.\xspace}
\catcode`\@=11
\def\@bibitem#1{%
\ifmc@bstsupport
  \mc@iftail{#1}%
    {;\newline\ignorespaces}%
    {\ifmc@first\else.\fi\orig@bibitem{#1}}
  \mc@firstfalse
\else
  \mc@iftail{#1}%
    {\ignorespaces}%
    {\orig@bibitem{#1}}%
\fi}%
\catcode`\@=12

\begin{table}[p]
\begin{center}
\begin{tabular}{| c || c | c | c | c | c | c | c | c |} \hline
LQ type & $\tilde S_{1/2}^L$ & $S_{1/2}^L$ & $S_{1/2}^R$ & $V_0^L$ & $V_0^R$ & $\tilde V_0^R$ & $V_1^L$ \\ \hline \hline
$\mu$-channel limit on $\MLQ (\gev)$ & 273 & 293 & 293 & 274 & 278 & 296 & 299 \\ \hline
$\tau$-channel limit on $\MLQ (\gev)$ & 270 & 291 & 291 & 271 & 276 & 294 & 298 \\ \hline
\end{tabular}
\caption{ \CL{95} lower limits on $\MLQ$ for $F=0$ LQs in the $\mu$- and the $\tau$-channels assuming $\lambda_{eq_1}=\lambda_{\ell q_{\beta}}=0.3$.}
\label{tab-LMF0}
\end{center}
\end{table}

\begin{table}[p]
\begin{center}
\begin{tabular}{| c || c | c | c | c | c | c | c | c |} \hline
LQ type & $S_{0}^L$ & $S_{0}^R$ & $\tilde S_{0}^R$ & $S_1^L$ & $V_{1/2}^L$ & $V_{1/2}^R$ & $\tilde V_{1/2}^R$ \\ \hline \hline
$\mu$-channel limit on $\MLQ (\gev)$ & 278 & 284 & 261 & 281 & 269 & 289 &
289 \\ \hline
$\tau$-channel limit on $\MLQ (\gev)$ & 275 & 281 & 257 & 278 & 265 & 287 &
286 \\ \hline
\end{tabular}
\caption{ \CL{95} lower limits on $\MLQ$ for $|F|=2$ LQs in the $\mu$- and the $\tau$-channels
assuming $\lambda_{eq_1}=\lambda_{\ell q_{\beta}}=0.3$.}
\label{tab-LMF2}
\end{center}
\end{table}

\begin{table}[p]
\begin{center}
\begin{tabular}{| c || c | c | c | c | c | c | c |} \hline
LQ type & $\tilde S_{1/2}^L$ & $S_{1/2}^L$ & $S_{1/2}^R$ & $V_0^L$/$V_0^R$ & $\tilde V_0^R$ & $V_1^L$ \\ \hline \hline
$\mu$-channel limit on $\lambda_{eq_1}\sqrt{\br{\mu q}}$  & 0.054 &
0.021 & 0.019 & 0.037 & 0.015 & 0.010 \\ \hline
$\tau$-channel limit on $\lambda_{eq_1}\sqrt{\br{\tau q}}$& 0.066 &
0.026 & 0.024 & 0.046  & 0.019 & 0.013 \\ \hline
\end{tabular}
\caption{\it \CL{95} upper limits on
  $\lambda_{eq_1}\sqrt{\br{\ell q}}$ for $F=0$ LQs
  with mass $\MLQ=250{\rm \gev}$ in the $\mu$- and the
  $\tau$-channels.}
\label{tab-COUPF0}
\end{center}
\end{table}
\begin{table}[p]
\begin{center}
\begin{tabular}{| c || c | c | c | c | c | c | c |} \hline
LQ type & $S_{0}^L$/$S_{0}^R$ & $\tilde S_{0}^R$ & $S_1^L$ & $V_{1/2}^L$ & $V_{1/2}^R$ & $\tilde V_{1/2}^R$ \\ \hline \hline
$\mu$-channel limit on $\lambda_{eq_1}\sqrt{\br{\mu q}}$  & 0.047 & 0.12 & 0.041 & 0.080
& 0.030 & 0.033 \\ \hline
$\tau$-channel limit on $\lambda_{eq_1}\sqrt{\br{\tau q}}$  & 0.058 & 0.15 & 0.049 & 0.10 & 0.038 & 0.042 \\ \hline
\end{tabular}
\caption{\it \CL{95} upper limits on
  $\lambda_{eq_1}\sqrt{\br{\ell q}}$ for $|F|=2$ LQs
  with mass $\MLQ=250{\rm \gev}$ in the $\mu$- and the $\tau$-channels.}
\label{tab-COUPF2}
\end{center}
\end{table}

\begin{table}
\footnotesize{
\begin{tabular}{|c||c|c|c|c|c|c|c|} \hline
\multicolumn{4}{|c}{} & \multicolumn{4}{c|}{} \\ 
\multicolumn{2}{|c}{\large{$e \rightarrow \mu$}}& 
\multicolumn{4}{c}{\normalsize{ZEUS $e^{\pm}p$ 94-00}} & 
\multicolumn{2}{c|}{\large{$F=0$}} \\ 
\multicolumn{4}{|c}{} & \multicolumn{4}{c|}{} \\ \hline
 & & & & & & & \\
\large{$\alpha \beta$} & \large{$S_{1/2}^L$}&\large{$S_{1/2}^R$}&\large{$\tilde{S}_{1/2}^L$}&\large{$V_0^L$}&\large{$V_0^R$}&\large{$\tilde{V}_0^R$}&\large{$V_1^L$}\\ 
&$e^- \bar{u}$&$e^-(\bar{u}+\bar{d})$&$e^-\bar{d}$&$e^- \bar{d}$&$e^-\bar{d}$&$e^-\bar{u}$&$e^-(\sqrt{2}\bar{u}+\bar{d})$\\ 
&$e^+u$&$e^+(u+d)$&$e^+d$&$e^+d$&$e^+d$&$e^+u$&$e^+(\sqrt{2}u+d)$\\ \hline \hline
   &$\mu N \rightarrow e N$ & $\mu N \rightarrow e N$ & $\mu N \rightarrow e N$ & $\mu N \rightarrow e N$ & $\mu N \rightarrow e N$ & $\mu N \rightarrow e N$ & $\mu N \rightarrow e N$ \\ 
1 1&$5.2 \times 10^{-5}$ & $2.6 \times 10^{-5}$ & $5.2 \times 10^{-5}$ & $2.6 \times 10^{-5}$ & $2.6 \times 10^{-5}$ & $2.6 \times 10^{-5}$ &  $0.8 \times 10^{-5}$\\     
                 & \boldmath $1.2$ & \boldmath $1.0$ & \boldmath $1.7$ & \boldmath $1.0$ & \boldmath $1.0$ & \boldmath $0.8$ & \boldmath $0.4$ \\ \hline
            & $D \rightarrow \mu \bar{e}$ & $K \rightarrow \mu \bar{e}$ & $K \rightarrow \mu \bar{e}$  & $K \rightarrow \mu \bar{e}$ & $K \rightarrow \mu \bar{e}$ & $D \rightarrow \mu \bar{e}$ & $K \rightarrow \mu \bar{e}$ \\ 
1 2&$2.4$&$2\times10^{-5}$&$2\times10^{-5}$&$1\times10^{-5}$&$1\times10^{-5}$&$1.2$&$1 \times 10^{-5}$\\     
                 & \fcolorbox{black}{yellow}{\boldmath $1.3$} & \boldmath $1.0$ & \boldmath $1.8$ &\boldmath $1.2$ & \boldmath $1.2$ & \fcolorbox{black}{yellow}{\boldmath $1.0$} & \boldmath $0.5$ \\ \hline
&  & $B \rightarrow \mu \bar{e}$ & $B \rightarrow \mu \bar{e}$ & $B \rightarrow \mu \bar{e}$ & $B \rightarrow \mu \bar{e}$ &  & $B \rightarrow \mu \bar{e}$ \\ 
  1 3       &  \boldmath $*$ & $0.4$& $0.4$  & $0.2$& $0.2$ & \boldmath $*$  & $0.2$  \\     
& \boldmath&\boldmath$1.8$&\boldmath$1.9$&\boldmath$1.5$&\boldmath$1.5$ & \boldmath & \boldmath $1.5$\\\hline
& $D \rightarrow \mu \bar{e}$ & $K \rightarrow \mu \bar{e}$ & $K \rightarrow \mu \bar{e}$  & $K \rightarrow \mu \bar{e}$ & $K \rightarrow \mu \bar{e}$ & $D \rightarrow \mu \bar{e}$ & $K \rightarrow \mu \bar{e}$ \\ 
2 1  & $2.4$ & $2 \times 10^{-5}$& $2 \times 10^{-5}$& $1\times 10^{-5}$& $1 \times 10^{-5}$&$1.2$&$1 \times 10^{-5}$\\     
& \boldmath $3.6$ & \boldmath $2.4$ & \boldmath $3.1$ & \boldmath $1.3$ & \boldmath $1.3$ & \boldmath $1.2$ & \boldmath $0.6$ \\ \hline
            & $\mu N \rightarrow e N$ & $\mu N \rightarrow e N$ & $\mu N \rightarrow e N$ & $\mu N \rightarrow e N$ & $\mu N \rightarrow e N$ & $\mu N \rightarrow e N$ & $\mu N \rightarrow e N$ \\ 
  2 2 & $9.2 \times 10^{-4}$& $1.3 \times 10^{-3}$& $3 \times 10^{-3}$ & $1.5 \times 10^{-3} $ & $ 1.5 \times 10^{-3}$ & $4.6 \times 10^{-4}$ &  $2.7 \times 10^{-4} $          \\     
                 & \boldmath $5.7$ & \boldmath $3.1$ & \boldmath $3.8$ & \boldmath $1.9$ & \boldmath $1.9$ & \boldmath $2.8$ & \boldmath $1.1$ \\ \hline
            &  & $B \rightarrow \bar{\mu} e K$  & $B \rightarrow \bar{\mu} e K$ & $B \rightarrow \bar{\mu} e K$ & $B \rightarrow \bar{\mu} e K$ & & $B \rightarrow \bar{\mu} e K$    \\ 
  2 3 & \boldmath $*$ & $0.3$ & $0.3$& $0.15$& $0.15$& \boldmath{$*$}& $0.15$\\     
& \boldmath & \boldmath $4.3$ & \boldmath $4.2$ & \boldmath $2.9$ & \boldmath $2.9$ & \boldmath & \boldmath $2.9$ \\ \hline
&  & $B \rightarrow \mu \bar{e}$  & $B \rightarrow \mu \bar{e}$ & $V_{ub}$ & $B \rightarrow \mu \bar{e}$ & & $V_{ub}$ \\ 
  3 1& \boldmath $*$ & $0.4$ & $0.4$ & $0.12$  & $0.2$  & \boldmath $*$ & $0.12$  \\     
& \boldmath & \boldmath $4.4$ & \boldmath $4.4$ & \boldmath $1.5$ & \boldmath $1.5$ & \boldmath & \boldmath $1.5$ \\ \hline
 &  & $B \rightarrow \bar{\mu} e K$ & $B \rightarrow \bar{\mu} e K$  & $B \rightarrow \bar{\mu} e K$ & $B \rightarrow \bar{\mu} e K$ & & $B \rightarrow \bar{\mu} e K$     \\ 
  3 2 & \boldmath $*$  & $0.3$& $0.3$& $0.15$ & $0.15$  & \boldmath $*$ & $0.15$\\     
& \boldmath & \boldmath $5.8$ & \boldmath $5.8$ & \boldmath $2.2$ & \boldmath $2.2$ & \boldmath & \boldmath $2.2$ \\ \hline
&  & $\mu N \rightarrow e N$ & $\mu N \rightarrow e N$ & $\mu N \rightarrow e N$ & $\mu N \rightarrow e N$ & & $\mu N \rightarrow e N$ \\ 
  3 3 & \boldmath $*$ &  $ 1.3 \times 10^{-3}$ & $3 \times 10^{-3} $   & $1.5 \times 10^{-3}$ & $1.5 \times 10^{-3}$ &  \boldmath $*$   &  $2.7 \times 10^{-4}$ \\     
& \boldmath & \boldmath $7.6$ & \boldmath $7.6$ & \boldmath $3.9$ & \boldmath $3.9$ & \boldmath & \boldmath $3.9$ \\ 
\hline
\end{tabular}
}

\caption{Limits at  $95\%$ C.L. on $\frac{\lambda_{eq_\alpha} \lambda_{\mu q_\beta}}{M^2_{LQ}}$ for $F=0$ LQs, in units of  $\tev^{-2}$. The first column indicates the quark generations coupling to $LQ - e$ and $LQ-\mu$, respectively. ZEUS results are reported in the third line (bold) of each cell. The low-energy process providing the most stringent constraint and the corresponding limit are shown in the first and second lines. The ZEUS limits are enclosed in a box if they are better than the low-energy constraints. The cases marked with * correspond to processes where the coupling to a $t$ quark is involved.}
\label{tab-HMF0MU}
\end{table}

%\vspace{-1.3cm}
\begin{table}
\footnotesize{
\begin{tabular}{|c||c|c|c|c|c|c|c|} \hline
\multicolumn{4}{|c}{} & \multicolumn{4}{c|}{} \\ 
\multicolumn{2}{|c}{\large{$e \rightarrow \mu$}}& 
\multicolumn{4}{c}{\normalsize{ZEUS $e^{\pm}p$ 94-00}} & 
\multicolumn{2}{c|}{\large{$|F|=2$}} \\ 
\multicolumn{4}{|c}{} & \multicolumn{4}{c|}{} \\ \hline
 & & & & & & & \\
\large{$\alpha \beta$} & \large{$S_0^L$}                    & \large{$S_0^R$}              
& \large{$\tilde{S}_0^R$}       & \large{$S_1^L$}          & \large{$V_{1/2}^L$}      & \large{$V_{1/2}^R$}      & \large{$\tilde{V}_{1/2}^L$} \\ 
 & $e^- u $ & $e^- u $ & $e^- d $ & $e^-(u+ \sqrt{2} d)$ & $e^- d $ & $e^-(u+d)$              & $e^- u $\\ 
& $e^+\bar{u} $ & $e^+ \bar{u} $& $e^+ \bar{d} $ & $e^+(\bar{u}+ \sqrt{2} \bar{d})$    & $e^+ \bar{d} $                & $e^+(\bar{u}+\bar{d})$              & $e^+\bar{u} $ \\ \hline \hline
& $\mu N \rightarrow e N$ & $\mu N \rightarrow e N$     & $\mu N \rightarrow e N$      & $\mu N \rightarrow e N$ & $\mu N \rightarrow e N$ & $\mu N \rightarrow e N$ & $\mu N \rightarrow e N$ \\ 
    1 1 & $5.2 \times 10^{-5}$& $5.2 \times 10^{-5}$        & $5.2 \times 10^{-5}$         & $1.7 \times 10^{-5}$    & $2.6 \times 10^{-5}$    & $1.3 \times 10^{-5}$    & $2.6 \times 10^{-5}$\\     
                 & \boldmath $1.6$ & \boldmath $1.6$ & \boldmath $2.1$ & \boldmath $0.9$ & \boldmath $0.9$ & \boldmath $0.5$ & \boldmath $0.6$ \\ 
\hline
& $K \rightarrow \pi \nu \bar{\nu}$ & $D \rightarrow \mu \bar{e}$ & $K \rightarrow \mu \bar{e}$  & $K \rightarrow \mu \bar{e}$ & $K \rightarrow \mu \bar{e}$ & $K \rightarrow \mu \bar{e}$&$D \rightarrow \mu \bar{e}$\\ 
1 2 & $10^{-3}$ & $2.4$& $2 \times 10^{-5}$& $1\times 10^{-5}$& $1 \times 10^{-5}$    & $1 \times 10^{-5}$    &  $1.2$          \\     
& \boldmath $2.5$ & \boldmath $2.5$ & \boldmath $2.6$ & \boldmath $1.2$ & \boldmath $1.6$ & \boldmath $1.2$ & \boldmath $1.8$ \\ \hline
            & &  & $B \rightarrow \mu \bar{e}$  & $V_{ub}$ & $B \rightarrow \mu \bar{e}$ & $B \rightarrow \mu \bar{e}$ &  \\ 
    1 3        &\boldmath $*$  &  \boldmath $*$ & $0.4$&$0.24$  & $0.2$& $0.2$&\boldmath $*$\\     
                 & \boldmath & \boldmath & \boldmath $2.9$ & \boldmath $1.4$ & \boldmath $2.2$ & \boldmath $2.2$ & \boldmath \\ \hline
&$K \rightarrow \pi \nu \bar{\nu}$ &$D \rightarrow \mu \bar{e}$ & $K \rightarrow \mu \bar{e}$  & $K \rightarrow \mu \bar{e}$ & $K \rightarrow \mu \bar{e}$ & $K \rightarrow \mu \bar{e}$ & $D \rightarrow \mu \bar{e}$ \\ 
2 1& $10^{-3}$& $2.4$& $2\times 10^{-5}$ & $1 \times 10^{-5}$   & $1\times 10^{-5}$ & $1 \times 10^{-5}$    &  $1.2$          \\   
 & \boldmath $2.1$ & \fcolorbox{black}{yellow}{\boldmath $2.1$} & \boldmath $2.5$ & \boldmath $1.1$ & \boldmath $0.9$ & \boldmath $0.5$ & \fcolorbox{black}{yellow}{\boldmath $0.6$} \\ \hline
& $\mu N \rightarrow e N$           & $\mu N \rightarrow e N$     & $\mu N \rightarrow e N$      & $\mu N \rightarrow e N$ & $\mu N \rightarrow e N$ & $\mu N \rightarrow e N$ & $\mu N \rightarrow e N$ \\ 
    2 2 & $9.2 \times 10^{-4}$ & $9.2 \times 10^{-4}$ & $3 \times 10^{-3}$ & $2.5 \times 10^{-3}$ & $1.5 \times 10^{-3} $ & $6.7 \times 10^{-4}$ &  $4.6 \times 10^{-4}$\\     
& \boldmath $5.7$ & \boldmath $5.7$ & \boldmath $3.8$ & \boldmath $1.8$ & \boldmath $1.9$ & \boldmath $1.6$ & \boldmath $2.8$ \\ \hline
            & &  & $B \rightarrow \bar{\mu} e K$  & $B \rightarrow \bar{\mu} e K$ & $B \rightarrow \bar{\mu} e K$ & $B \rightarrow \bar{\mu} e K$ &    \\ 
    2 3 & \boldmath$*$& \boldmath $*$ & $0.3$& $0.15$& $0.15$   & $0.15$ &  \boldmath $*$    \\     
& \boldmath & \boldmath & \boldmath $4.4$ & \boldmath $2.2$ &\boldmath $2.9$ & \boldmath $2.9$ & \boldmath \\\hline
&  &  & $B \rightarrow \mu \bar{e}$  & $B \rightarrow \mu \bar{e}$ & $B \rightarrow \mu \bar{e}$ & $B \rightarrow \mu \bar{e}$ &  \\ 
    3 1 & \boldmath$*$  & \boldmath $*$ & $0.4$&$0.4$  & $0.2$                     & $0.2$  & \boldmath $*$ \\     
                 & \boldmath & \boldmath & \boldmath $3.1$ & \boldmath $1.5$ & \boldmath $0.9$ & \boldmath $0.9$ & \boldmath\\ \hline
            & &  & $B \rightarrow \bar{\mu} e K$  & $B \rightarrow \bar{\mu} e K$ & $B \rightarrow \bar{\mu} e K$ & $B \rightarrow \bar{\mu} e K$ &    \\ 
    3 2        &       \boldmath $*$             & \boldmath $*$  & $0.3$& $0.15$                       & $0.15$ & $0.15$ & \boldmath $*$     \\     
                 & \boldmath & \boldmath & \boldmath $5.9$ & \boldmath $3.0$ & \boldmath $2.2$ & \boldmath $2.2$ & \boldmath \\ \hline
& & & $\mu N \rightarrow e N$ & $\mu N \rightarrow e N$ & $\mu N \rightarrow e N$ & $\mu N \rightarrow e N$ &  \\ 
    3 3 &   \boldmath $*$   &  \boldmath $*$  & $3 \times 10^{-3}$ & $2.5 \times 10^{-3}$              & $1.5\times 10^{-3}$  & $6.7 \times 10^{-4}$ &  \boldmath $*$         \\     
& \boldmath & \boldmath & \boldmath $7.7$ & \boldmath $3.9$ & \boldmath $4.0$ & \boldmath $4.0$ & \boldmath \\ \hline
\end{tabular}
}

\caption{Limits at $95\%$ C.L. on $\frac{\lambda_{eq_\alpha} \lambda_{\mu q_\beta}}{M^2_{LQ}}$ for $F=2$ LQs, in units of $\tev^{-2}$. The first column indicates the quark generations coupling to $LQ - e$ and $LQ-\mu$, respectively. ZEUS results are reported in the third line (bold) of each cell. The low-energy process providing the most stringent constraint and the corresponding limit are shown in the first and second lines. The ZEUS limits are enclosed in a box if they are better than the low-energy constraints. The cases marked with * correspond to processes where the coupling to a $t$ quark is involved.}
\label{tab-HMF2MU}
\end{table}

%\vspace{-1.3cm}
\begin{table}
\footnotesize{

\begin{tabular}{|c||c|c|c|c|c|c|c|} \hline
\multicolumn{4}{|c}{} & \multicolumn{4}{c|}{} \\ 
\multicolumn{2}{|c}{\large{$e \rightarrow \tau$}}&
\multicolumn{4}{c}{\normalsize{ZEUS $e^{\pm}p$ 94-00}} & 
\multicolumn{2}{c|}{\large{$F=0$}} \\ 
\multicolumn{4}{|c}{} & \multicolumn{4}{c|}{} \\ \hline
 & & & & & & & \\
\large{$\alpha \beta$} & \large{$S_{1/2}^L$} & \large{$S_{1/2}^R$}      & \large{$\tilde{S}_{1/2}^L$} & \large{$V_0^L$} & \large{$V_0^R$} & \large{$\tilde{V}_0^R$} & \large{$V_1^L$} \\ 
                 & $e^- \bar{u} $     & $e^-(\bar{u}+\bar{d}) $ & $e^- \bar{d} $             & $e^- \bar{d} $ & $e^- \bar{d} $ & $e^- \bar{u}$          & $e^-(\sqrt{2}\bar{u}+\bar{d})$\\ 
                 & $e^+u $     & $e^+(u+d) $ & $e^+d $             & $e^+d $ & $e^+d $ & $e^+u$          & $e^+(\sqrt{2}u+d)$\\ \hline \hline
            & $\tau \rightarrow \pi e$ & $\tau \rightarrow \pi e$ & $\tau \rightarrow \pi e$ & $\tau \rightarrow \pi e$ & $\tau \rightarrow \pi e$ & $\tau \rightarrow \pi e$ & $\tau \rightarrow \pi e$ \\ 
  1 1          & $0.4$ & $0.2$ & $0.4$ & $0.2$ & $0.2$ & $0.2$ &  $0.06$\\     
                 & \boldmath $1.8$ & \boldmath $1.5$ & \boldmath $2.7$ & \boldmath $1.7$ & \boldmath $1.7$ & \boldmath $1.3$ & \boldmath $0.6$ \\ \hline
            && $\tau \rightarrow K e$ & $K \rightarrow \pi \nu \bar{\nu}$ & $\tau \rightarrow K e$ & $\tau \rightarrow K e$ & & $K \rightarrow \pi \nu \bar{\nu}$ \\ 
  1 2          && $6.3$  & $5.8 \times 10^{-4}$  & $3.2$  & $3.2$ & &  $1.5 \times 10^{-4}$\\     
                 & \fcolorbox{black}{yellow}{\boldmath $1.9$} & \fcolorbox{black}{yellow}{\boldmath $1.6$} & \boldmath $2.9$ & \fcolorbox{black}{yellow}{\boldmath $2.1$} & \fcolorbox{black}{yellow}{\boldmath $2.1$} & \fcolorbox{black}{yellow}{\boldmath $1.6$} & \boldmath $0.8$ \\ \hline
         &  & $B \rightarrow \tau \bar{e}$ & $B \rightarrow \tau \bar{e}$  & $B \rightarrow \tau \bar{e}$ & $B \rightarrow \tau \bar{e}$ &  & $B \rightarrow \tau \bar{e}$ \\ 
  1 3       &  \boldmath $*$ & $0.3$  & $0.3$ & $0.13$  & $0.13$ & \boldmath $*$  & $0.13$  \\     
            & \boldmath & \boldmath $3.2$ & \boldmath $3.3$ & \boldmath $2.6$ & \boldmath $2.6$ & \boldmath & \boldmath $2.6$ \\ \hline
            & & $\tau \rightarrow K e$ & $K \rightarrow \pi \nu \bar{\nu}$ & $\tau \rightarrow K e$ & $\tau \rightarrow K e$ & & $K \rightarrow \pi \nu \bar{\nu}$ \\ 
  2 1          && $6.3$ & $5.8 \times 10^{-4}$ & $3.2$ & $3.2$ & & $1.5 \times 10^{-4}$ \\     
               & \fcolorbox{black}{yellow}{\boldmath $6.0$} &  \fcolorbox{black}{yellow}{\boldmath $4.1$} & \boldmath $5.2$ &\fcolorbox{black}{yellow}{ \boldmath $2.3$} & \fcolorbox{black}{yellow}{\boldmath $2.3$} & \fcolorbox{black}{yellow}{\boldmath $2.1$} & \boldmath $0.9$ \\ \hline
            & $\tau \rightarrow 3e$ & $\tau \rightarrow 3e$  & $\tau \rightarrow 3e$ & $\tau \rightarrow 3e$ & $\tau \rightarrow 3e$ & $\tau \rightarrow 3e$ & $\tau \rightarrow 3e$ \\ 
  2 2          & $5$  & $8$ & $17$ & $9$ & $9$ & $3$ &  $1.6$   \\     
                 & \boldmath $10$ & \fcolorbox{black}{yellow}{\boldmath $5.6$} & \fcolorbox{black}{yellow}{\boldmath $6.5$} & \fcolorbox{black}{yellow}{\boldmath $3.4$} & \fcolorbox{black}{yellow}{\boldmath $3.4$} & \boldmath $5.5$ & \boldmath $2.1$ \\ \hline
 &  & $B \rightarrow \tau \bar{e} X$  & $B \rightarrow \tau \bar{e} X$ & $B \rightarrow \tau \bar{e} X$ & $B \rightarrow \tau \bar{e} X$ & & $B \rightarrow \tau \bar{e} X$    \\ 
  2 3          & \boldmath $*$ & $14$ & $14$ & $7.2$ & $7.2$ & \boldmath{$*$}& $7.2$   \\     
& \boldmath & \fcolorbox{black}{yellow}{\boldmath $8.1$} & \fcolorbox{black}{yellow}{\boldmath $7.8$}& \fcolorbox{black}{yellow}{\boldmath $5.5$} & \fcolorbox{black}{yellow}{\boldmath $5.5$} & \boldmath & \fcolorbox{black}{yellow}{\boldmath $5.5$}\\ \hline
            &  & $B \rightarrow \tau \bar{e}$ & $B \rightarrow \tau \bar{e}$  & $V_{ub}$ & $B \rightarrow \tau \bar{e}$ &  & $V_{ub}$ \\ 
  3 1          &  \boldmath $*$ & $0.3$  & $0.3$ & $0.12$  & $0.13$  & \boldmath $*$ & $0.12$  \\     
& \boldmath & \boldmath $7.8$ & \boldmath $7.2$ & \boldmath $2.5$ & \boldmath $2.5$ & \boldmath & \boldmath $2.5$ \\ \hline
            &  & $B \rightarrow \tau \bar{e} X$  & $B \rightarrow \tau \bar{e} X$ & $B \rightarrow \tau \bar{e} X$ & $B \rightarrow \tau \bar{e} X$ & & $B \rightarrow \tau \bar{e} X$    \\ 
  3 2          & \boldmath $*$  & $14$  & $14$ & $7.2$ & $7.2$ & \boldmath $*$ & $7.2$ \\     
   & \boldmath & \fcolorbox{black}{yellow}{\boldmath $11$} & \fcolorbox{black}{yellow}{\boldmath $10$} & \fcolorbox{black}{yellow}{\boldmath $4.2$} & \fcolorbox{black}{yellow}{\boldmath $4.2$} & \boldmath & \fcolorbox{black}{yellow}{\boldmath $4.2$} \\ \hline
            &               &  $\tau \rightarrow 3e$ &  $\tau \rightarrow 3e$ & $\tau \rightarrow 3e$ & $\tau \rightarrow 3e$ &  & $\tau \rightarrow 3e$\\ 
  3 3        & \boldmath $*$ &  $8$ & $17$   & $9$ & $9$ &  \boldmath $*$   &  $1.6$ \\     
                 & \boldmath & \boldmath $15$ & \fcolorbox{black}{yellow}{\boldmath $14$} & \fcolorbox{black}{yellow}{\boldmath $8.1$} &\fcolorbox{black}{yellow}{ \boldmath $8.1$} & \boldmath & \boldmath $8.1$ \\ \hline
 \end{tabular}
}

\caption{Limits at $95\%$ C.L. on $\frac{\lambda_{eq_\alpha} \lambda_{\tau q_\beta}}{M^2_{LQ}}$ for $F=0$ LQs, in units of $\tev^{-2}$. The first column indicates the quark generations coupling to $LQ - e$ and $LQ-\tau$, respectively. ZEUS results are reported in the third line (bold) of each cell. The low-energy process providing the most stringent constraint and the corresponding limit are shown in the first and second lines. The ZEUS limits are enclosed in a box if they are better than the low-energy constraints. The cases marked with * correspond to processes where the coupling to a $t$ quark is involved.}

\label{tab-HMF0TAU}
\end{table}
%

%\vspace{-1.3cm}
\begin{table}
\footnotesize{
\begin{tabular}{|c||c|c|c|c|c|c|c|} \hline
\multicolumn{4}{|c}{} & \multicolumn{4}{c|}{} \\
\multicolumn{2}{|c}{\large{$e \rightarrow \tau$}} & 
\multicolumn{4}{c}{\normalsize{ZEUS $e^{\pm}p$ 94-00}} & 
\multicolumn{2}{c|}{\large{$|F|=2$}} \\ 
\multicolumn{4}{|c}{} & \multicolumn{4}{c|}{} \\ \hline
 & & & & & & & \\
\large{$\alpha \beta$} & \large{$S_0^L$}                    & \large{$S_0^R$}              & \large{$\tilde{S}_0^R$}       & \large{$S_1^L$}          & \large{$V_{1/2}^L$}      & \large{$V_{1/2}^R$}      & \large{$\tilde{V}_{1/2}^L$} \\ 
                 & $e^- u $                          & $e^- u $                    & $e^- d $                     & $e^-(u+ \sqrt{2} d)$    & $e^- d $                & $e^-(u+d)$              & $e^- u $\\ 
                 & $e^+\bar{u} $                          & $e^+ \bar{u} $& $e^+ \bar{d} $                     & $e^+(\bar{u}+ \sqrt{2} \bar{d})$    & $e^+ \bar{d} $                & $e^+(\bar{u}+\bar{d})$              & $e^+\bar{u} $ \\ \hline \hline
                 & $G_F$           & $\tau \rightarrow \pi e$     & $\tau \rightarrow \pi e$      & $\tau \rightarrow \pi e$ & $\tau \rightarrow \pi e$ & $\tau \rightarrow \pi e$ & $\tau  \rightarrow \pi e$ \\ 
    1 1          & 0.3 & 0.4 & 0.4 & 0.1 & 0.2 & 0.1 & 0.2\\      
                 & \boldmath $2.5$ & \boldmath $2.5$ & \boldmath $3.5$ & \boldmath $1.4$ & \boldmath $1.4$ & \boldmath $0.8$ & \boldmath $1.0$ \\ \hline
                 & $K \rightarrow \pi \nu \bar{\nu}$ & & $\tau \rightarrow K e$  & $K \rightarrow \pi \nu \bar{\nu}$ & $K \rightarrow \pi \nu \bar{\nu}$ & $\tau \rightarrow K  e$ & \\ 
    1 2  & $5.8 \times 10^{-4}$ & & $6.3$ & $2.9 \times 10^{-4}$ & $2.9 \times 10^{-4}$ & $3.2$ &\\     
         & \boldmath $4.0$ & \fcolorbox{black}{yellow}{\boldmath $4.0$} &\fcolorbox{black}{yellow}{ \boldmath $4.4$} & \boldmath $1.9$ & \boldmath $2.8$ & \fcolorbox{black}{yellow}{\boldmath $2.0$} & \fcolorbox{black}{yellow}{\boldmath 
$3.1$} \\ \hline
                 &  &  & $B \rightarrow \tau \bar{e}$  & $V_{ub}$ & $B \rightarrow \tau \bar{e}$ & $B \rightarrow \tau \bar{e}$ &  \\ 
    1 3 & \boldmath $*$&  \boldmath $*$ & $0.3$ & $0.12$  & $0.13$ & $0.13$  &  \boldmath $*$\\     
        & \boldmath & \boldmath & \boldmath $5.1$ & \boldmath $2.6$ & \boldmath $4.0$ & \boldmath $4.0$ & \boldmath \\ \hline
            & $K \rightarrow \pi \nu \bar{\nu}$ & & $\tau \rightarrow K e$  & $K \rightarrow \pi \nu \bar{\nu}$ & $K \rightarrow \pi \nu \bar{\nu}$ & $\tau \rightarrow K  e$ & \\ 
    2 1        & $5.8 \times 10^{-4}$ & & $6.3$ & $2.9 \times 10^{-4}$ & $2.9 \times 10^{-4}$  & $3.2$ &\\   
              & \boldmath $3.2$ & \fcolorbox{black}{yellow}{\boldmath $3.2$} & \fcolorbox{black}{yellow}{\boldmath $4.3$} & \boldmath $1.8$ & \boldmath $1.4$ & \fcolorbox{black}{yellow}{\boldmath $0.8$} & \fcolorbox{black}{yellow}{\boldmath $1.0$} \\ \hline
            & $\tau \rightarrow 3e$ & $\tau \rightarrow 3e$  & $\tau \rightarrow 3e$  & $\tau \rightarrow 3e$ & $\tau \rightarrow 3e$ & $\tau \rightarrow 3e$ & $\tau \rightarrow 3e$ \\ 
    2 2        & $5$ & $5$ & $17$ & $14$ & $9$ & $4$ &  $3$\\     
               & \boldmath $10$ & \boldmath $10$ & \fcolorbox{black}{yellow}{\boldmath $6.5$} & \fcolorbox{black}{yellow}{\boldmath $3.2$} & \fcolorbox{black}{yellow}{\boldmath $3.5$} & \fcolorbox{black}{yellow}{\boldmath $2.8$} & \boldmath $5.1$ \\ \hline
            & &  &  $B \rightarrow \tau \bar{e} X$  & $B \rightarrow \tau \bar{e} X$ & $B \rightarrow \tau \bar{e} X $ & $B \rightarrow \tau \bar{e} X$ &    \\ 
    2 3        & \boldmath $*$ & \boldmath $*$  & $14$ & $7.2$ & $7.2$ & $7.2$ & \boldmath $*$   \\     
               & \boldmath & \boldmath & \fcolorbox{black}{yellow}{\boldmath $8.3$} & \fcolorbox{black}{yellow}{\boldmath $4.1$} & \fcolorbox{black}{yellow}{\boldmath $5.4$} & \fcolorbox{black}{yellow}{\boldmath $5.4$} & \boldmath \\ \hline
            & &  & $B \rightarrow \tau \bar{e}$  & $B \rightarrow \tau \bar{e}$ & $B 
\rightarrow \tau \bar{e}$ & $B \rightarrow \tau \bar{e}$ &  \\ 
    3 1        & \boldmath $*$  & \boldmath $*$ & $0.3$  & $0.13$  & $0.13$ & $0.13$  & \boldmath $*$ \\ 
                 & \boldmath & \boldmath & \boldmath $5.3$ & 
\boldmath $2.7$ & \boldmath $1.6$ & \boldmath $1.6$ & \boldmath \\ \hline    
            & & &  $B \rightarrow \tau \bar{e} X$  & $B \rightarrow \tau \bar{e} X$ & $B \rightarrow \tau \bar{e} X $ & $B \rightarrow \tau \bar{e} X$ &    \\ 
    3 2        & \boldmath $*$  &\boldmath $*$  & $14$  & $7.2$ & $7.2$ & $7.2$ & \boldmath $*$     \\     
                 & \boldmath & \boldmath & \fcolorbox{black}{yellow}{\boldmath $11$} & \fcolorbox{black}{yellow}{\boldmath $5.5$} & \fcolorbox{black}{yellow}{\boldmath $4.1$} & \fcolorbox{black}{yellow}{\boldmath $4.1$} & \boldmath \\ \hline
            &            &      & $\tau \rightarrow 3e$  & $\tau \rightarrow 3e$ & $\tau \rightarrow 3e$ & $\tau \rightarrow 3e$ &          \\ 
    3 3        &   \boldmath $*$   &  \boldmath $*$  & $17$  & $14$ & $9$  & $4$ &  \boldmath $*$    \\     
                 & \boldmath & \boldmath & \fcolorbox{black}{yellow}{\boldmath $15$} & \fcolorbox{black}{yellow}{\boldmath $7.6$} & \fcolorbox{black}{yellow}{\boldmath$7.6$} & \boldmath $7.6$ & \boldmath \\ 
\hline

\end{tabular}
}

\caption{Limits at $95\%$ C.L. on $\frac{\lambda_{eq_\alpha} \lambda_{\tau q_\beta}}{M^2_{LQ}}$ for $F=2$ LQs, in units of $\tev^{-2}$. The first column indicates the quark generations coupling to $LQ - e$ and $LQ-\tau$, respectively. ZEUS results are reported in the third line (bold) of each cell. The low-energy process providing the most stringent constraint and the corresponding limit are shown in the first and second lines. The ZEUS limits are enclosed in a box if they are better than the low-energy constraints. The cases marked with * correspond to processes where the coupling to a $t$ quark is involved.}
\label{tab-HMF2TAU}
\end{table}


\begin{figure}[p]
\vfill
\begin{center}
\epsfig{file=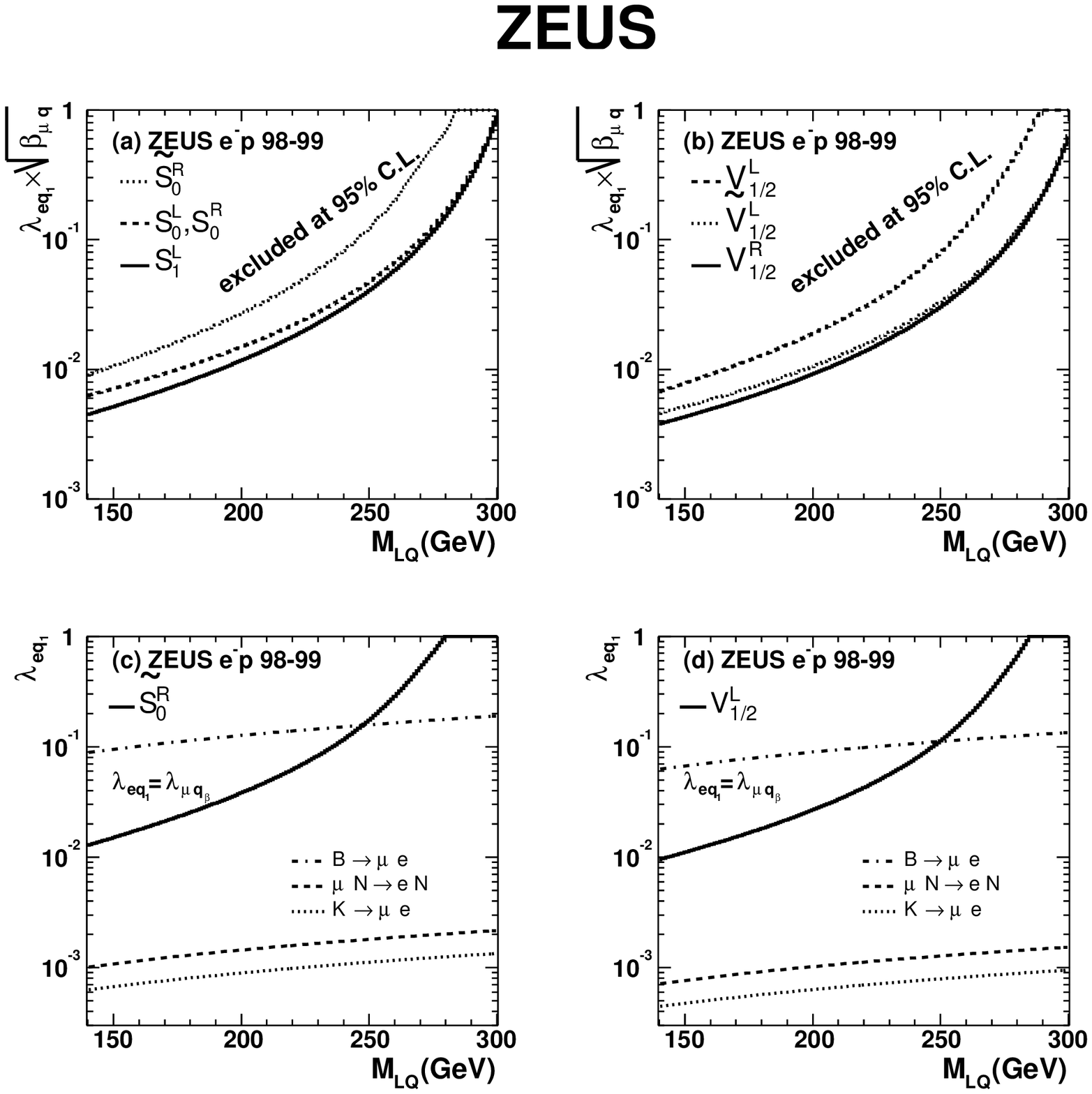, width=\textwidth}
\end{center}
\caption{Limits for $F=2$ low-mass LQs in the $\mu$ channel obtained from $e^{-}p$ collisions. The upper plots show \CL{95} limits on $\lambda_{eq_1}\times\sqrt{\beta_{\mu q}}$ for (a) scalar and (b) vector LQs. In the lower plots, ZEUS limits on $\lambda_{eq_1}$ for a representative (c) scalar and (d) vector LQ are compared to the indirect constraints  from low-energy experiments~\protect\cite{zfp:c61:613,*pr:d62:055009,*pr:d66:010001,*Herz:2002gq,*Aubert:2003pc,*Yusa:2004gm,*prl:93:241802}, assuming $\lambda_{eq_1}=\lambda_{\mu q_{\beta}}$.
}
\label{fig-mu_f2}
\vfill
\end{figure}

\begin{figure}[p]
\vfill
\begin{center}
\epsfig{file=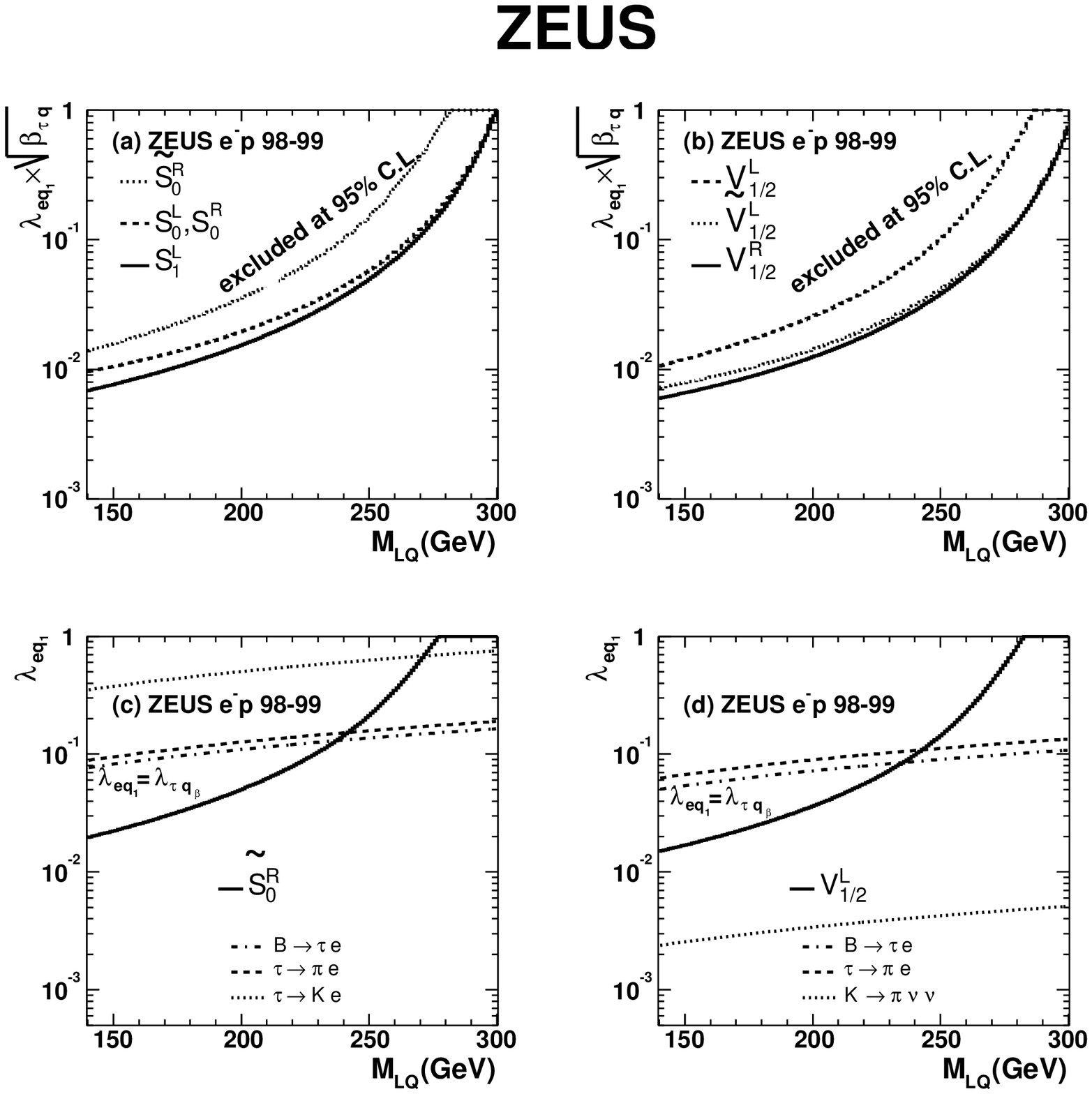, width=\textwidth}
\end{center}
\caption{Limits for $F=2$ low-mass LQs in the $\tau$ channel obtained from $e^{-}p$ collisions. The upper plots show \CL{95} limits on $\lambda_{eq_1}\times\sqrt{\beta_{\tau q}}$ for (a) scalar and (b) vector LQs. In the lower plots, ZEUS limits on $\lambda_{eq_1}$ for a representative (c) scalar and (d) vector LQ are compared to the indirect constraints  from low-energy experiments~\protect\cite{zfp:c61:613,*pr:d62:055009,*pr:d66:010001,*Herz:2002gq,*Aubert:2003pc,*Yusa:2004gm,*prl:93:241802}, assuming $\lambda_{eq_1}=\lambda_{\tau q_{\beta}}$.
}
\label{fig-tau_f2}
\vfill
\end{figure}

\end{document}